# Power transport efficiency during O-X-B 2nd harmonic electron cyclotron heating in a helicon linear plasma device[1]

J. F. Caneses Marin[a,b], C. L. Lau[a], R.H. Goulding[a], T. Bigelow[a], T. Biewer[a], J. B. O. Caughman[a], J. Rapp[a]

[a]Oak Ridge National Laboratory, Fusion Energy Division, Oak Ridge, TN 37831, USA

[b]E-mail: canesesmarjf@ornl.gov

**Abstract:**
The principal objective of this work is to report on the power coupled to a tungsten target in the Proto-MPEX device during oblique injection of a microwave beam (< 70 kW at 28 GHz) into a high-power (~100 kW at 13.56 MHz) over-dense ($n_e > 1 \times 10^{19}$ m$^{-3}$) deuterium helicon plasma column. The experimental setup, electron heating system, electron heating scheme, and IR thermographic diagnostic for quantifying the power transport is described in detail. It is demonstrated that the power transported to the target can be effectively controlled by adjusting the magnetic field profile. Using this method, heat fluxes up to 22 MWm$^{-2}$ and power transport efficiencies in the range of 17-20% have been achieved using 70 kW of microwave power. It is observed that most of the heat flux is confined to a narrow region at the plasma periphery. Ray-tracing calculations are presented which indicate that the power is coupled to the plasma electrons via an O-X-B mode conversion process. Calculations indicate that the microwave power is absorbed in a single pass at the plasma periphery via collisions and in the over-dense region via 2nd harmonic cyclotron resonance of the electron Bernstein wave. The impact of these results is discussed in the context of MPEX.

## 1 Introduction

The Material Plasma Exposure eXperiment (MPEX) is a new linear divertor simulator under construction at Oak Ridge National Laboratory which aims to address key Plasma Material Interaction (PMI) science and technology questions for future fusion reactors [1]. Such questions include hydrogen retention in Plasma-Facing Components (PFCs) exposed to reactor-relevant ion fluences, end of life testing of PFCs, increasing Technology-Readiness-Level (TRL) of PFC power exhaust solutions, etc. To achieve this, MPEX will expose a-priori neutron-irradiated samples under steady-state (up to $10^6$ seconds) fusion-relevant divertor plasmas with independent control of electron and ion temperatures via the application of auxiliary heating systems [1].

Present day linear divertor simulators for PMI research make use of a variety of plasma source technologies to create low-temperature high-density plasmas [2]. Such plasma sources include DC arc [3], high-pressure cascaded arc [4] and helicon sources [5], [6]. These plasma source technologies can expose materials to ion fluences 1 – 2 orders of magnitude higher than those produced in present-day toroidal fusion experiments [2]. An overview of existing and planned PMI linear plasma devices is provided in reference [2]. Generally, these plasma source technologies are very efficient at ionizing neutral gas but are restricted to low electron temperatures (1-6 eV) and independent control of electron and ion temperature in these linear configurations has not been demonstrated. Notable exceptions include ion heating with Radio-Frequency (RF) power in the VASIMR device [7] and electron heating with 3 GHz microwaves in a low-density afterglow discharge [8].

[1] This manuscript has been authored by UT-Battelle, LLC, under contract DE-AC05-00OR22725 with the US Department of Energy (DOE). The publisher acknowledges the US government license to provide public access under the DOE Public Access Plan (http://energy.gov/downloads/doe-public-access-plan)

To overcome this limitation, pioneering research on electron [9]–[11] and ion heating [12]–[14] in a linear plasma configuration has been carried at ORNL to support the MPEX design [1]. The basic strategy taken to increase electron and ion temperature is to couple auxiliary heating power at suitable cyclotron resonance locations present in an already formed low-temperature high-density plasma column created by a helicon plasma source. A key feature of helicon plasma sources is the production of plasmas with very high ionization fraction (up to 90% [5]), which translates to excellent neutral gas management and differential pumping [15] in the plasma heating regions. With baffles and neutral gas skimmers, the low neutral gas content achievable in the plasma heating regions (~0.01 Pa) makes helicon plasma sources exceptionally compatible with the application of auxiliary heating to increase electron and ion temperatures. In-depth description of the various heating schemes developed in MPEX is presented in references [9], [10], [13], [14].

In this work, we focus on the heat fluxes coupled to a tungsten target and the associated parallel power transport efficiency during $2^{nd}$ harmonic Electron Bernstein Wave (EBW) heating experiments (< 70 kW at 28 GHz) in low-temperature high-density deuterium helicon plasmas (100-120 kW at 13.56 MHz) produced in the Proto-MPEX device. Moreover, we revisit the following question: What is the optimal magnetic field profile to maximize the power transport to the target? This question has been investigated in previous numerical [16] and experimental work [10]; however, there are two main reasons for revisiting this matter: (1) significant improvements in electron heating have been produced leading to heat loads to the target between 3 to 10 times higher than those previously reported (see references [1], [10], [16]); (2) the IR diagnostic used for power transport measurements in references [1], [10], [16] did not capture the entire plasma footprint and thus underestimated the total amount of power coupled to the target. The IR diagnostic used in the present work overcomes this limitation and allows, for the first time, calculation of absolute power transport efficiencies during electron heating.

This paper is structured as follows: The experimental setup and the electron heating chamber are described in detail in section 2.1 and 2.2 respectively. The IR-based diagnostic setup employed to quantify the power transport is presented in detail in section 2.3. The heat flux reconstruction procedure and numerical analysis is described in section 3. The O-X-B electron heating scheme employed in this work is presented in detail in section 4. In the results (section 5), the scaling of the power transport with applied heating power and magnetic field profile is presented. In section 6, ray tracing calculations are presented to gain insight into the electron heating scheme. Finally, in section 7 the results are discussed in the context of MPEX.

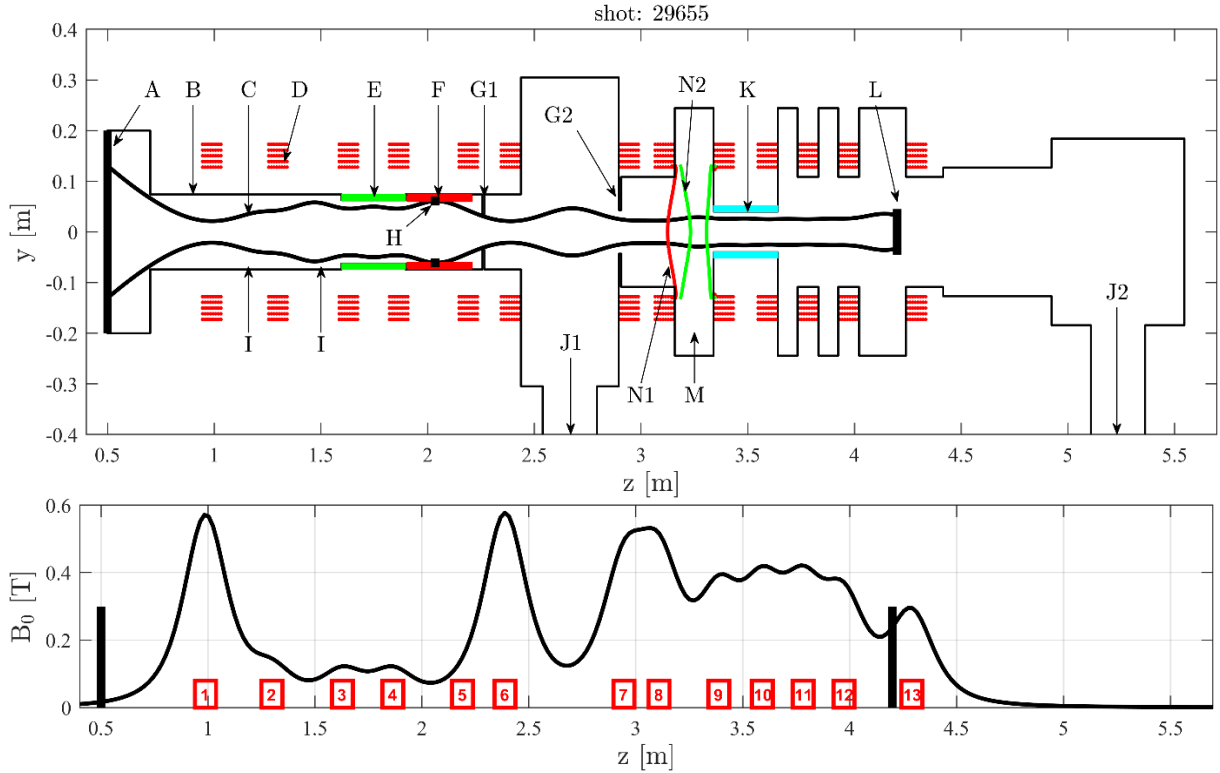

Figure 1, (top) Schematic of Proto-MPEX. (bottom) Magnetic field along the axis of the device. Labelled items represent the following: (A) "dump" plate, (B) vacuum boundary, (C) Last Uninterrupted Flux Surface (LUFS), (D) electromagnet, (E) AlN helicon dielectric window, (F) Stainless steel cylindrical limiter, (G1) and (G2) skimmers, (H) plasma strike point on limiter, (I) neutral gas injection location, (J1) and (J2) location of turbo-molecular pumps, (K) alumina ICH dielectric window, (L) tungsten target plate, (M) 28 GHz injection location, (N1 and N2) 2$^{nd}$ and 3$^{rd}$ harmonic electron cyclotron resonance layers respectively.

## 2 Experimental setup

The electron heating experiments herein described have been performed in the Proto-MPEX helicon-based linear plasma device. In this work, the presentation of the experimental setup is divided into the following parts: (1) helicon plasma source, (2) electron heating chamber and (3) the target-IR thermographic diagnostic. Each part is now presented in detail.

### 2.1 Helicon plasma source

A schematic of the Proto-MPEX device is presented in Figure 1. An example of the magnetic field along the axis of the device is also provided. The most important components are labelled A through N and are described in the caption. The helicon plasma source consists of an Aluminum Nitride (AlN) dielectric window (Item E) 30 cm in length, 6 mm in thickness and 12.5 cm inner diameter. The helicon dielectric window is surrounded by a quarter-turn right-handed helical antenna 22 cm in length and driven at 13.56 MHz by a 200 kW RF transmitter and a matching network.

#### 2.1.1 Neutral gas management and differential pumping

Deuterium gas is injected at locations labelled I in Figure 1 at a typical rate of 0.7 – 1.5 SLM (Standard Liters per Minute) based on the applied RF power. Two skimmers (items G1 and G2) are used for neutral gas management and production of differential pumping in the electron heating chamber (item M). Skimmer G1 decreases neutral gas diffusion from the plasma source (~0.5 to 1 Pa) into the heating vacuum chambers

(~ 0.01 Pa). Skimmer G2 provides another barrier directly upstream of the electron heating chamber. Moreover, the Ion Cyclotron Heating (ICH) dielectric window (item K) whose primary purpose is to couple RF power to the plasma ions [14] also serves as a conduction limiting element. The ICH window contributes to the differential pumping between the target chamber (~0.5 to 1 Pa) and the electron heating chamber. These conduction limiting elements (G1, G2 and K) have been demonstrated to sustain neutral gas pressures in the heating chambers down to 0.01 Pa [15]. Effective differential pumping in the heating chambers have been demonstrated to be *critical* for efficient electron heating as reported in reference [17].

Neutral gas is extracted from the device at two locations (labelled J1 and J2) with turbo-molecular pumps rated at 3.8 $m^3s^{-1}$ and 2.5 $m^3s^{-1}$. The largest pump (J1) is located closest to the electron heating chamber. However, it has been demonstrated that the plasma provides a pumping capacity much greater than the J1 turbo-molecular pump and in fact excellent differential pumping and plasma performance can be sustained just with the J2 target chamber pump [5].

### 2.1.2 Plasma density and temperature measurements

A Double Langmuir Probe (DLP) [18] , positioned 5 cm upstream of the target plate, is used to measure the radial profiles of plasma density and electron temperature during the electron heating experiments. The DLP consists of two molybdenum current collecting tips 2 mm in length and 0.254 mm in diameter separated by approximately 2 mm. The DLP circuit is magnetically coupled to an ambipolar power supply via a 1:1 isolation transformer and driven with a triangular wave at 200 Hz with amplitudes between 20-80 V peak to peak based on the operating conditions. The collected current is measured with a Pearson current transformer model 4100. The voltage difference between the DLP tips is measured across a shunt resistor (2.2 kOhms) and another Pearson isolation transformer model 4100. The DLP output signals are digitized at a rate of 5 kHz and stored in an MDSplus database.

### 2.1.3 Magnetic field profile

The background magnetic field is produced using 13 coils (labelled in Figure 1) and powered with 5 different DC power supplies (PS). Each PS has a specific function and is connected to group of coils as follows: (1) "Helicon mirror" coils: 1 and 6, (2) "Helicon source" coils: 3 and 4, (3) "Helicon shaping" coil 2, (4) "Electron cyclotron" coils: 7 and 8 and (5) "Ion cyclotron" coils: 9 to 13.

The shape of the magnetic field along the axis of the device is determined by various requirements. The helicon plasma source requires a magnetic field less than 0.2 Tesla for optimal operation. Magnetic mirrors on either side of the helicon source are used to improve plasma production as shown in the following references [6], [19]. In addition, the magnetic field at the electron heating chamber needs to be close to 0.5 Tesla to provide a 2$^{nd}$ harmonic electron cyclotron resonance layer during electron heating experiments with 28 GHz microwave power.

### 2.1.4 Axial and radial boundaries

The axial boundaries of the plasma are defined by the "dump" and "target" plates (items A and L respectively). The target consists of a tungsten plate (90 × 90 $mm^2$) 0.75 mm in thickness which has been surface-treated to increases its emissivity up to a value of 0.3 in the IR wavelength of interest (see Figure 6 and section 2.3). The "dump" plate consists of a stainless steel plate of 0.4 m diameter and 1.5 mm in thickness. Item C represents the so-called "Last Uninterrupted Flux Surface" (LUFS) [20] and defines the radial extent of the plasma. The LUFS is defined by the point of contact of the plasma at the strike point (item H), which for the experiments reported in this paper, occurs on a dedicated stainless-steel cylindrical limiter (item F). Details on the use of this limiter and its impact on the helicon dielectric window heating is presented in reference [20].

The LUFS is calculated using an accurate model of the magnetic coils and the dimensions of the device. Each magnetic coil is modelled as a collection of current filaments using realistic dimensions, number of turns and layers. The LUFS is calculated using the vacuum magnetic field generated by the coils. Given the plasma densities and temperatures typically measured in Proto-MPEX ($n_e < 1 \times 10^{20}$ m$^{-3}$, $T_e \approx T_i < 5$ eV) and the magnetic field ($B \sim [0.1 - 1]$ T), the plasma kinetic pressure ($< 160$ Pa) is much less than the magnetic pressure (4 kPa to 400 kPa); hence, the diamagnetic effect of the plasma on the magnetic field and LUFS can be neglected. Reference [20, Fig. 4] compares the calculated LUFS and the experimentally observed LUFS based on plasma density and temperature measurements.

## 2.2 Electron heating chamber

A 1:1 scale side-view and sectional-view of the electron heating chamber is presented in Figure 2. Items labelled A through G represent the following: (A) plane used for creating the sectional-view A-A, (B) plasma edge (LUFS), (C) 2$^{nd}$ harmonic electron cyclotron harmonic surface closest to the 28 GHz injection location, (D) corrugated circular waveguide employed to inject the 28 GHz microwave radiation into an ellipsoidal stainless steel reflector (item G), (E) vacuum boundary. Item F represents the outline of the tungsten target as seen from the target and looking towards the plasma source. The tungsten target has a 45 degrees azimuthal rotation about the main axis of the device given by the "z" coordinate (see Figure 7). The ellipsoidal reflector (item G) focuses the microwave beam and makes the wavefronts conformal to the plasma column's cylindrical shape to improve power coupling. GENRAY-C numerical results presented in reference [9] indicate that beam focusing significantly improves power coupling to the plasma column. The ellipsoidal reflector (item G) and 28 GHz waveguide (item D) are presented in the isometric view of Figure 3. The ellipsoidal reflector enables the injection of the microwaves at an angle of *incidence* of 30 degrees (see Figure 1 left). The *incidence* angle is the angle between the beam path and the plane normal to the plasma surface as defined in optics.

The microwave power injected into the electron heating chamber is measured with a calibrated diode detector mounted on the waveguide system and the plasma-induced power incident on the target plate is measured with the IR imagining system. However, absolutely-calibrated measurements of radiated power, microwave power absorbed in the plasma and power leaked into other regions of the vacuum chamber are not available.

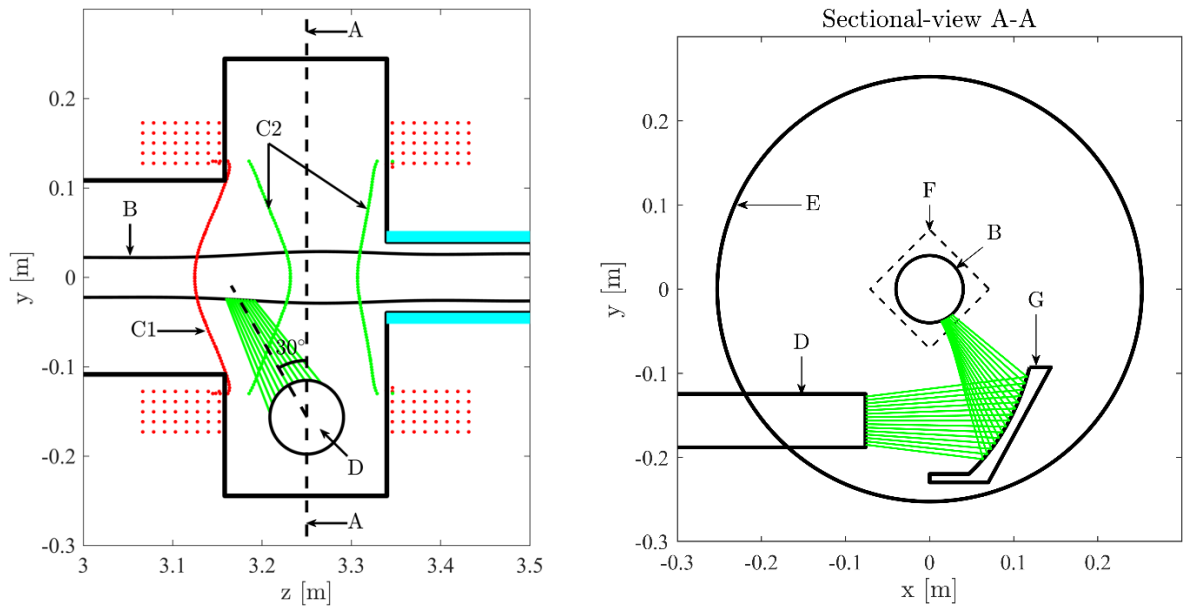

Figure 2, 1:1 scale (left) side-view and (right) sectional-view of the electron heating chamber. Items labelled represent the following: (A) sectional plane, (B) plasma boundary (LUFS), (C1 and C2) 2nd and 3rd harmonic electron cyclotron resonance surfaces, (D) corrugated circular waveguide where the 28 GHz is injected and aimed at an ellipsoidal reflector, (E) vacuum boundary, (F) outline of tungsten target as seen from target-side of device, (G) ellipsoidal reflector. The green lines represent rays which approximate the trajectory of the 28 GHz microwave radiation incident on the plasma column.

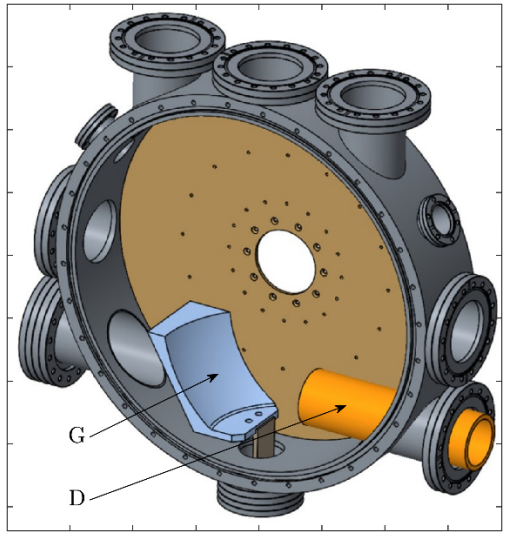

Figure 3, Isometric view of the electron heating chamber. Item G and D represent the ellipsoidal reflector and waveguide used to inject 28 GHz microwave radiation respectively.

## 2.3 Target-IR diagnostic setup

To infer the power transported to the target via plasma flux, an IR imaging camera which views the "back-side" of the tungsten target is used as illustrated in Figure 4. In all the data presented in this work, the plasma column intersects the target plate at a normal angle of incidence; moreover, the IR camera views the "back-side" of the target also at a normal angle of incidence. The diagnostic technique relies on collecting IR radiant flux from the target to infer its surface temperature. The principles of IR thermography relevant to this work are well described in references [20], [21]. Using the temporal and spatial distribution of the surface temperature, the plasma-induced heat flux is inferred using the appropriate physics models and numerical methods (see section 3).

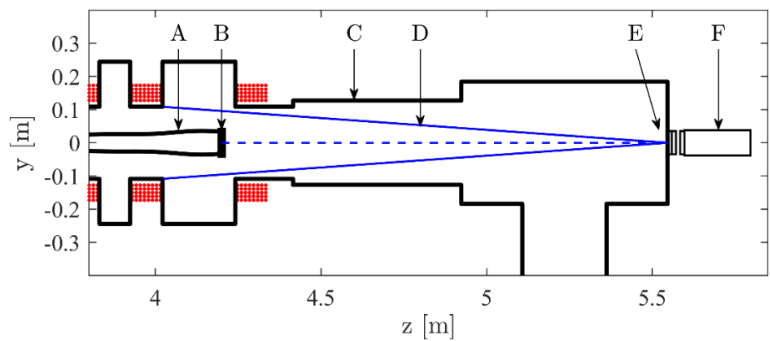

Figure 4, 1:1 scale schematic of the target plate and IR thermographic diagnostic setup. Labelled items represent the following: (A) plasma boundary (LUFS), (B) tungsten target plate, (C) vacuum boundary, (D) light rays representing the Field Of View (FOV) of the IR camera, (E) viewing port with ZnSe window and (F) FLIR A655 IR camera.

The IR camera employed in this study is a FLIR A655 with a spectral band of $7 - 14$ μm and a frame rate of 100 Hz. Surface temperatures in the range of -40 to 650 degrees C can be measured with the FLIR A655 IR camera and has an option to measure temperatures up to 2000 degrees C. The ZnSe window provides a transmission factor of 0.7 in this IR wavelength band. In this setup, the plasma flux is incident on the so-called "front-side" of the target while the IR radiant flux is collected from the so-called "back-side" surface. This strategy eliminates the uncertainty caused by IR emissivity changes due to plasma-induced surface modifications as depicted in Figure 5. The composition and the mechanism which forms these surface features is currently under investigation and will be the subject of another paper.

Moreover, the dimensions of the tungsten target are large enough to intercept that entire plasma column and the IR camera FOV is large enough to collect the entire plasma-induced IR radiant flux emitted by the "back-side" of the target. This allows the estimation of the total power coupled to the target by the plasma during electron heating experiments; hence, provides a way to calculate the efficiency by which the electron heating scheme couples power to the target plate. This diagnostic setup is an improvement over the one described in reference [22] which had a partial view of the "front-side" of the target, was susceptible to the effect of surface modifications and did not capture the full plasma footprint. However, using a "back-side" imaging method for the target comes at a cost: estimating the heat flux incident on the "front-side" based on the IR radiant flux emitted from the "back-side" requires an inverse solution. This is further described in section 3.

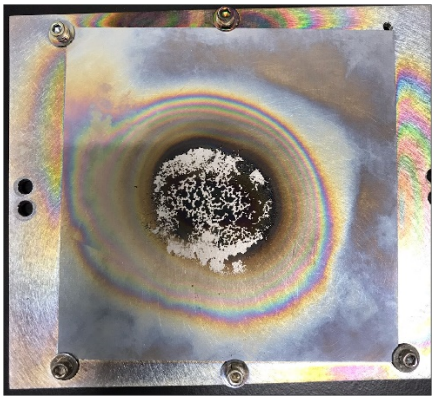

Figure 5, Example illustrating the surface modification and deposits formed on the "front-side" (plasma-exposed) surface of the tungsten target after multiple months of plasma operation. These features can introduce important changes to the surface IR emissivity and errors in the thermographic analysis. No IR emissivity changes were observed on the "back-side" surface of the target.

IR thermography of metallic surfaces can be challenging due to their high reflectivity. For this reason, the tungsten target used in the present study was surface-treated to increase its IR emissivity. After surface treatment, the surface roughness was measured to be about $R_a = 0.82$ μm and the IR emissivity was experimentally determined to be about 0.3 for the viewing angle and FOV shown in Figure 4. The experimental data used to determine the emissivity of the surface-treated tungsten target is presented in Figure 6. Based in the linearity of the IR-TC data, in the limited temperature range of 300 to 415 K, the emissivity of 0.3 can be safely extrapolated to about 600 K. To extend this range to higher temperatures, we make use of the experimental data presented in reference [23]. In this reference, the IR emissivity of tungsten samples is measured (in vacuum) in the temperature range of 470 K to 1120 K. The measurements indicate that the IR emissivity has a weak linear dependency with temperature with a slope of 1e-4 per kelvin in this temperature range. In the present work, the temperature range of interest lies between 300 K and 900 K; hence, based on reference [23] and the data from Figure 6, the IR emissivity used in the analysis is taken to be $0.3 \pm 0.03$. The uncertainty in the emissivity is included in the heat flux measurement by calculating the heat flux using the lower and upper bound of the emissivity (0.27 and 0.33). This process is used to quantify the error bars presented in the heat flux measurements.

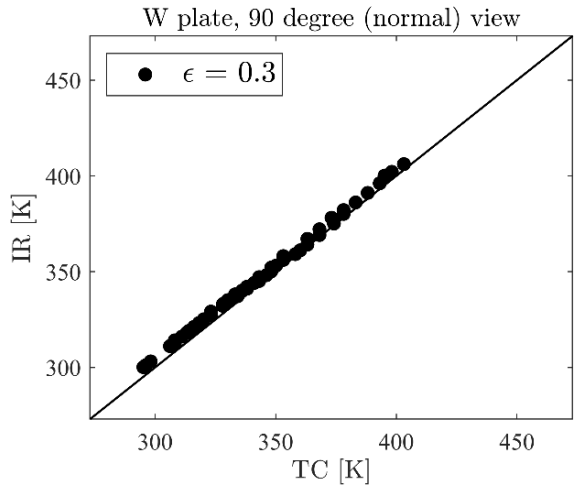

Figure 6, Infrared (IR) and thermocouple (TC) measurements used to determine the surface emissivity of the surface-treated tungsten target plate for normal incidence view (90 degrees). The surface roughness was measured to be about $R_a = 0.82$ μm. IR measurements were taken with a FLIR A655 IR camera with a spectral band of 7 to 14 μm. The surface-treated tungsten target was uniformly heated, and its temperature was recorded over a temperature range of 300 K to 420 K using the IR camera and a thermocouple (TC).

The tungsten target is positioned normal to the main axis of the device ("z" coordinate). The IR camera views the "back" surface of the target at a normal angle of incidence (see Figure 4); moreover, the plasma column intersects the tungsten target also at a normal angle of incidence. However, due to vacuum port availability and the mounting method, the target is positioned with a 45 degree azimuthal rotation about the main axis of the device ("z" coordinate) as shown in Figure 7. The target holder is held in place by a stainless-steel rod positioned at a 45 degree angle. Due to this tilt angle about the "z" coordinate, a local coordinate system $(\hat{x}_*, \hat{y}_*)$ is used to describe the target plate geometry and is subsequently used in this manuscript for the presentation of the inferred heat fluxes.

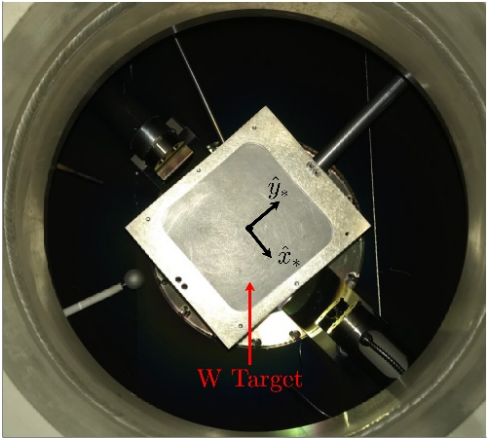

Figure 7, the tungsten target as seen by the IR camera. The unit vectors $\hat{x}_*$ and $\hat{y}_*$ represent the coordinate system local to the target plate and are used throughout this manuscript. IR camera collects the IR radiant flux originating from the surface shown ("back-side") while the plasma interacts with the side of the plate not shown ("front-side").

# 3   Heat flux reconstruction and numerical analysis

Reconstructing the "front-side" heat flux distribution based on the "back-side" surface temperature distribution requires solving the heat diffusion equation in a so-called "inverse" manner; more specifically, given the temperature distribution (the effect), we wish to determine the heat flux boundary condition (the cause). Problems of this type are known as "inverse" problems, are considered ill-posed and their solution may require advanced numerical methods [24]–[28]. The standard way of solving "inverse" problems is to pose them as optimization problems which minimize a cost function.

Formulating the cost function requires solving the direct or "forward" problem for a given iteration of the unknown boundary condition. In this context, "forward" refers to calculating the effect (temperature distribution) given the cause (heat flux boundary conditions). Hence, solving an inverse problem via minimization of a cost function requires a fast and accurate method for solving the forward problem.

## 3.1   Solution to the forward problem

Solving the heat diffusion problem in a "forward" manner typically involves specifying the heat flux boundary conditions and then calculating the resulting temperature distribution. In the present context, the geometry of the problem is illustrated in Figure 24. It consists of insulating boundary conditions on all faces of the target plate except the front-side where a time-dependent surface heat flux boundary condition is specified.

The dimensionless problem is described by Eq. 10 in Appendix 1: Forward problem. Only by considering the material properties as constants, this problem be solved via the Green's function method. Such solution is presented in  Eq. 15 in Appendix 1: Forward problem. The ability to express the solution as a closed-form expression significantly simplifies the solution to the inverse problem. In cases where closed-form solutions are not available and numerical solutions are needed, the so-called Alifanov iterative regularization method has been used, for example see references [25], [26].

It is important to note that while tungsten has a temperature dependent heat conductivity (see figure 7b in reference [29]), in the context of this work, selecting the mean thermal conductivity in the temperature range of interest (300 to 900 K) yields acceptable errors as demonstrated Figure 25a.

## 3.2 Validation of the inverse problem

Validating the solution to the inverse problem has been carried using the following steps: (1) Create a synthetic heat flux distribution $q(x_*, y_*, t)$ for the "front-side" surface, (2) solve the forward problem using the synthetic heat flux distribution $q(x_*, y_*, t)$ in a commercial FEM software package which includes *temperature-dependent* material properties. (3) Extract the time-dependent "back-side" surface temperature distribution from the FEM solution and use it as the input to an inverse problem with *constant* material properties. (4) Reconstruct the synthetic "front-side" heat flux distribution by solving the inverse problem. (5) Compare the synthetic and reconstructed "front-side" heat flux distributions for various plate thicknesses.

The validation process for the heat flux inverse reconstruction used in this work is presented in Appendix 2: Validating the solution to the inverse problem. It is shown that the inverse reconstruction process is well suited for the present experimental conditions based on a 0.75 mm thick tungsten target.

# 4 Electron heating scheme

The electron heating scheme reported in this paper makes use of 28 GHz microwave radiation. The O-mode cutoff density at this frequency is approximately $n_e = 1 \times 10^{19}$ m$^{-3}$; hence, injection of 28 GHz microwave radiation at a normal angle of incidence cannot access the core of the over-dense ($\omega_{pe} > \omega$) high-density plasmas created in Proto-MPEX ($n_e = [0.3 - 1] \times 10^{20}$ m$^{-3}$).

To overcome this limitation, an electron heating scheme which excites electron Bernstein waves (EBWs) or B-modes via mode-conversion is employed. EBWs are radially-propagating electrostatic hot-plasma waves which have no density cut-off and can be absorbed via *resonant* and *non-resonant* mechanisms [30]. EBWs propagate *almost* perpendicularly to the magnetic field hence they develop a finite parallel wave number. Due to their high refractive index, EBWs are strongly absorbed *resonantly* at doppler-shifted harmonic resonant surfaces [31]. Moreover, due to their electrostatic nature they are also strongly absorbed *non-resonantly* via collisions with neutrals [32] and/or Coulomb collisions [9]. This can be understood by noting that electrostatic waves carry a significant fraction of their energy in coherent particle motion which can be scattered via collisions. EBWs have been studied in stellarators [30], [33] and spherical tokamaks [31], [34] for heating and current drive. However, EBW physics and heating in linear plasma configurations has not received as much attention. Some examples include references [9], [10], [35], [36]. A more detailed description of the O-X-B conversion process applied to electron heating in linear configurations can be found in reference [9].

Since EBWs do not propagate in vacuum, their production relies on mode-conversion processes. There are two main heating schemes capable of exciting B modes using externally-launched power: (1) X-B and (2) O-X-B mode conversion. In the X-B process, an X-mode polarized microwave beam is injected at a normal angle of incidence to the magnetized plasma. If the density gradient is steep enough, the fast X-mode can tunnel through both the O-mode cutoff and the Upper Hybrid Resonance (UHR) layer to couple to the slow X-mode which then excites B modes at the UHR layer. Direct experimental observation of this process in a linear plasma configuration is presented in reference [36].

In the O-X-B process, an O-mode polarized microwave beam is injected *obliquely* to the magnetic field. When the *angle of incidence* and/or the parallel refractive index ($n_\parallel$) is given by Eq. 1 [9], [37], the O-mode cutoff is *coincident* with the slow X-mode cutoff (see Appendix 3) and very efficient O to X mode conversion is produced [34], [37]. The O-X mode conversion process can be observed clearly in full-wave calculations presented in [38] and consistent with ray-tracing theory. The newly excited slow X-mode propagates radially *outwards* towards the lower density region where the Upper-Hybrid Resonance (UHR) layer (Eq. 3) is located. At the UHR layer, the slow X-mode becomes increasingly electrostatic and its perpendicular wavenumber matches that of the EBW and full mode conversion from X to B mode occurs [9], [39]. The resulting Bernstein wave propagates radially *inwards* towards the higher density region. The Bernstein wave is strongly absorbed at nearby doppler-shifted electron cyclotron harmonic resonant surfaces. In addition, the Bernstein wave, being electrostatic in nature, is readily absorbed via collisions; hence, leading to both *resonant* and *non-resonant* wave absorption [30]. It is worth pointing out that the optimality of the angle of incidence for O to X mode conversion presented in Eq. 1 has been clearly demonstrated with a full-wave solution in reference [38].

$$\sin\theta = n_\parallel = \sqrt{\frac{Y}{1+Y}} \quad \text{where} \quad Y = \frac{\omega_{ce}}{\omega} \qquad \text{Eq. 1}$$

In the present experiments, the O-X-B mode conversion process is employed. The 28 GHz microwave beam is focused by an ellipsoidal reflector and injected with an O-mode polarization at an incident angle of about 30 degrees (see Figure 2). The magnetic field in the electron heating region is close to 0.5 T and with a microwave frequency of 28 GHz, the optimal angle of incidence based on Eq. 1 is approximately 35 degrees.

The O-X-B mode conversion process for $\omega_{ce}/\omega_{RF} > 0.5$ is illustrated in Figure 8 where $\omega_{RF}$ is the wave's frequency, $\omega_{ce}$ is the electron cyclotron frequency, $\hat{r}$ is the normalized radial coordinate of the plasma which becomes 1 at the 28 GHz O-mode cutoff density $n_e^O$; $\hat{n}$ is the plasma density profile normalized by $n_e^O$ ($\hat{n} = n_e/n_e^O$) and $\hat{z}$ is the coordinate along the length of the plasma column. The vertical dashed line represents the 2$^{nd}$ harmonic cyclotron resonance layer where the EBWs can be resonantly absorbed. The magnetic field is axially varying and close to 0.5 Tesla. The horizontal red line is the location where the microwave frequency matches the UHR frequency, and the black horizontal line is the O-mode cutoff layer. The upper hybrid frequency is given by Eq. 2 and the plasma density at which the UHR layer occurs $n_e^{UH}$ is given by Eq. 3, where $n_e^O$ is the O-mode cutoff density.

$$\omega_{UH}^2 = \omega_{pe}^2 + \omega_{ce}^2 \qquad \text{Eq. 2}$$

$$n_e^{UH} = n_e^O \left(1 - \left(\frac{\omega_{ce}}{\omega_{RF}}\right)^2\right) \quad \text{where} \quad n_e^O = \frac{m_e \epsilon_0}{e^2} \omega_{RF}^2 \qquad \text{Eq. 3}$$

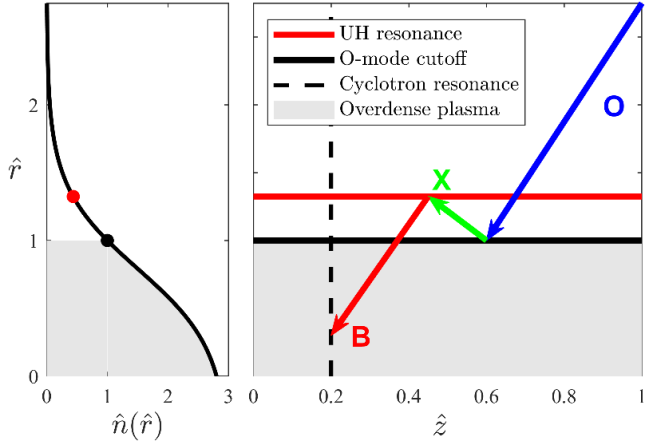

Figure 8, (left) plasma density profile normalized to the 28 GHz O-mode cutoff density $n_e^O$ where $\hat{n} = n_e/n_e^0$. Black circle represents the location of the O-mode cutoff density and red circle the location of the Upper Hybrid (UH) resonance layer. (right) r-z plasma slice illustrating the trajectory of O-mode microwave radiation and the associated mode-conversion processes involved. Grey shaded areas represent over-dense plasma regions ($\omega < \omega_{pe}$). A ratio $\omega_{ce}/\omega > 0.5$ is assumed.

# 5 Experimental results
## 5.1 Maximizing parallel power transport to the target

In this section, we revisit the following question: What is the optimal magnetic field profile to maximize the power transport to the target? This question has been investigated in previous numerical [16] and experimental work [10]; however, there are two main reasons for revisiting this matter: (1) significant improvements in electron heating have been produced leading to heat loads to the target between 3 to 10 times higher than those previously reported (see references [1], [10], [16]); (2) the IR diagnostic used for power transport measurements in references [1], [10], [16] did not capture the entire plasma footprint and thus underestimated the total amount of power coupled to the target. The IR diagnostic used in the present work overcomes this limitation and allows, for the first time, calculation of absolute power transport efficiencies during electron heating.

Electron heating experiments were performed using the setup described in section 2. Microwave radiation at 28 GHz and constant power of 70 kW was injected into deuterium helicon plasmas produced with 100 kW of RF power at 13.56 MHz. The relative timing between the RF and microwave power is illustrated in Figure 9. The RF pulse is approximately 500 ms in duration while the microwave pulse is 70 ms. The microwave pulse is injected approximately 300 ms after the initiation of the helicon plasma. During this time, the helicon plasma reaches steady-state conditions and the neutral gas pressure in the heating region has reached a stationary value of approximately 0.01 Pa (see [15, Fig. 12], [20, Fig. 15]). This ensures that the plasma column in the electron heating region has reached its highest ionization fraction ($> 0.9$) [5] and neutral gas content is low enough for efficient electron heating.

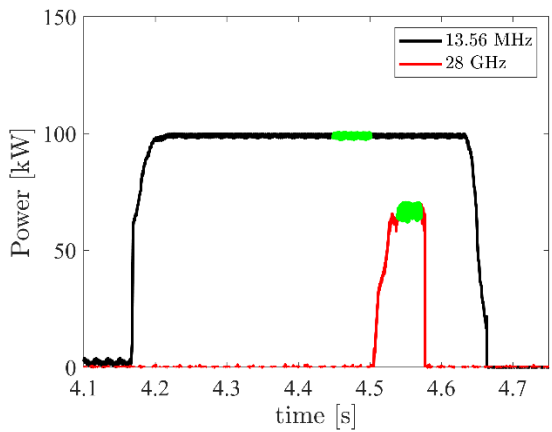

Figure 9, Typical time evolution of the applied RF (13.56 MHz) and microwave (28 GHz) power measured during electron heating experiments. The green line corresponds to the time interval used to calculate the time-averaged plasma density and temperature measurements presented in Figure 17.

The magnetic field profiles investigated during this experiment are presented in Figure 10. The location of the electron heating region (Figure 2 and Figure 3), where the 28 GHz power is injected, is represented by the grey region. The magnetic field strength corresponding to the $2^{nd}$ harmonic of 28 GHz is 0.5 T. Each magnetic field profile shown in Figure 10 has a 0.5 T $2^{nd}$ harmonic resonant surface within the electron heating region and is the location where the EBWs are expected to be *resonantly* absorbed.

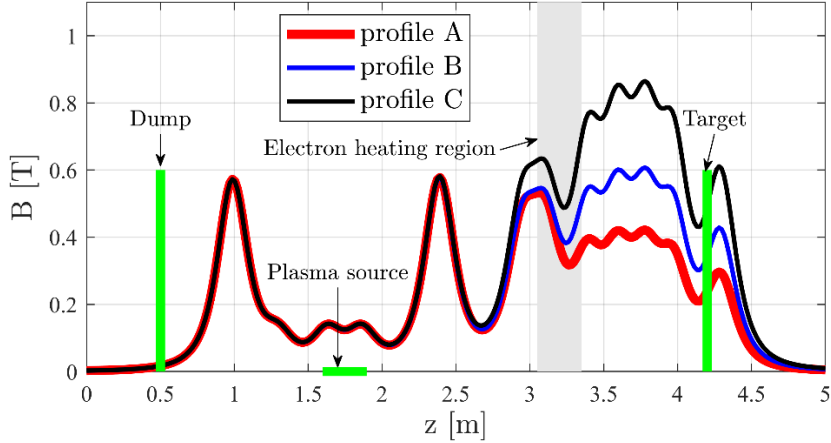

Figure 10, Magnetic field scenarios used to determine optimal magnetic field profile. Grey region illustrates the position of the 28 GHz microwave chamber (Figure 2).

### 5.1.1 2-D Reconstructed Target Heat Fluxes

Using the IR diagnostic described in section 2.3, the two-dimensional (2-D) plasma heat flux incident on the front-side of the tungsten target is reconstructed via an inverse method based on the IR radiant flux emitted by the back-side surface.

The 2-D reconstructed heat fluxes incident on the tungsten target during the application of the 28 GHz microwaves for all three magnetic field profiles (Figure 10) are presented in Figure 11. To reduce the shot noise, these heat flux distributions were time-averaged over 3 time steps. A mean emissivity of 0.3 was used in the analysis. Error bars due to the uncertainty in the emissivity ($\epsilon = 0.3 \pm 0.03$) can be found in Figure 12. The dotted circles represent the location of the LUFS based on magnetic flux calculations. The green arrows illustrate the location where the 28 GHz microwave radiation is injected by the ellipsoidal reflector relative to the plasma column. The green contour lines enclose regions with heat fluxes greater than 2.5 MWm$^{-2}$. In all the cases presented, the heat flux distribution can be described by two contributions: (1) a "peaked" and (2) a diffuse "halo" contribution that surrounds the LUFS. The "peaked" heat flux contribution is positioned directly at the location where the ellipsoidal reflector injects the 28 GHz microwave and is the location of highest heat fluxes up to ~12 MWm$^{-2}$ for the magnetic profile A (Figure 12 (left)). On the other hand, the "halo" contribution is evenly distributed all around the plasma column. This type of "halo" heat flux distribution has been previously observed in electron heating experiments in Proto-MPEX where the ellipsoidal reflector was not used [11].

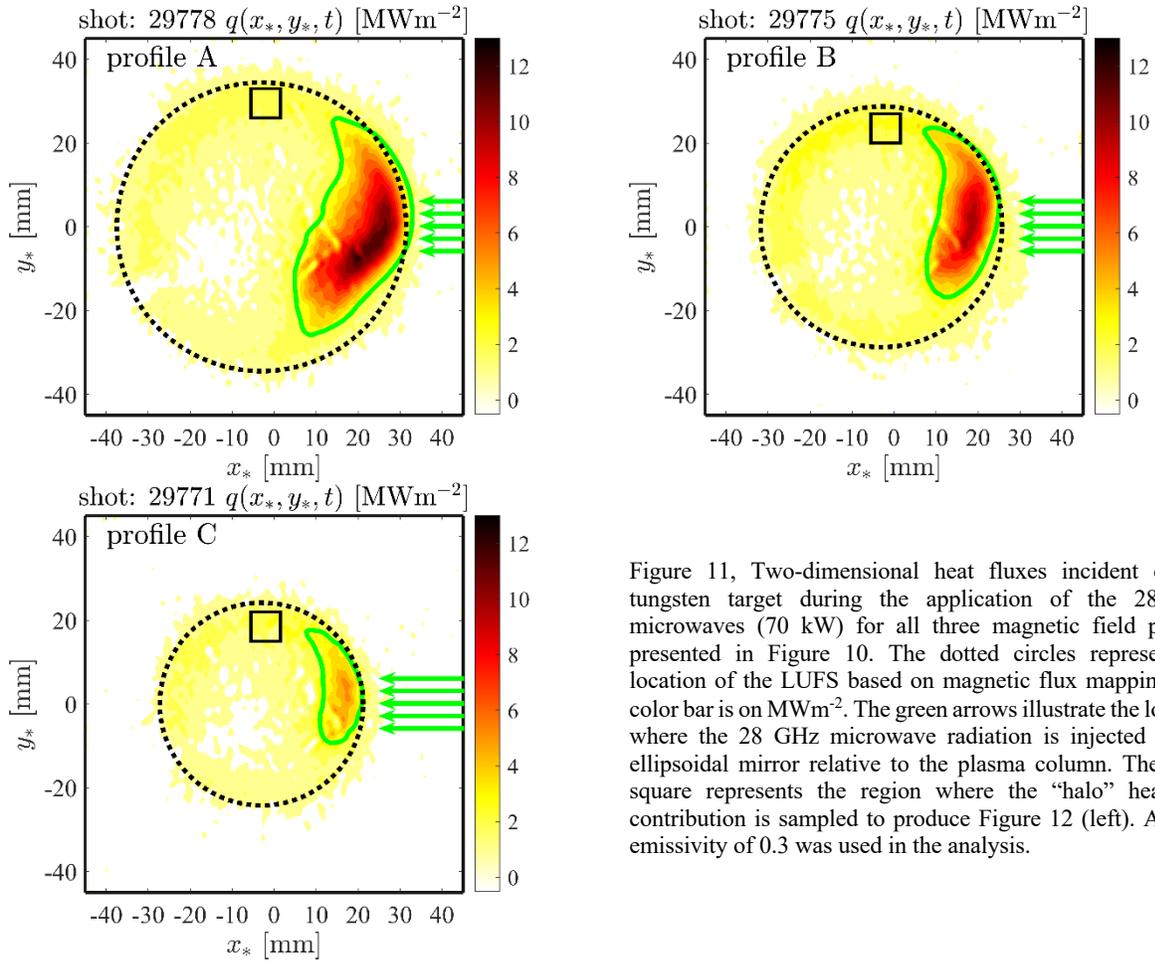

Figure 11, Two-dimensional heat fluxes incident on the tungsten target during the application of the 28 GHz microwaves (70 kW) for all three magnetic field profiles presented in Figure 10. The dotted circles represent the location of the LUFS based on magnetic flux mapping. The color bar is on MWm$^{-2}$. The green arrows illustrate the location where the 28 GHz microwave radiation is injected by the ellipsoidal mirror relative to the plasma column. The black square represents the region where the "halo" heat flux contribution is sampled to produce Figure 12 (left). A mean emissivity of 0.3 was used in the analysis.

Figure 12 (left) presents the scaling of the "peaked" and "halo" heat flux contributions as a function of the peak magnetic field near z = 3.8 m while keeping the 28 GHz microwave power constant at 70 kW. The heat flux is obtained from the data presented in Figure 11. Error bars are associated with the uncertainty in the emissivity ($\epsilon = 0.3 \pm 0.03$). The "halo" contribution is obtained from the region bounded by the black square shown in Figure 11. The "peaked" contribution is obtained from the maximum heat flux observed at the "hot spot" near the 28 GHz injection zone. The labels "A" to "C" in Figure 12 represent the magnetic field profile associated with these measurements. The data indicates that the "halo" heat flux contribution is *independent* of the magnetic field profile and has a constant value of approximately 2.5 MWm$^{-2}$ at this 28 GHz microwave power level. On the other hand, the "peaked" heat flux contribution scales inversely proportional to the peak magnetic field near z = 3.8 m and attains a global maximum value of about 12 MWm$^{-2}$ when using magnetic field profile A. Using magnetic field configuration "C", the "peaked" heat flux drops to about 5 MWm$^{-2}$. This demonstrates that for a fixed 28 GHz microwave power level, the contribution of the "peaked" heat flux coupled to the target can be controlled by the choice of the magnetic field profile. The "halo" contribution; however, is independent of the magnetic field profile. This suggests that the "peak" and "halo" heat flux distributions originate from *different* absorption mechanisms.

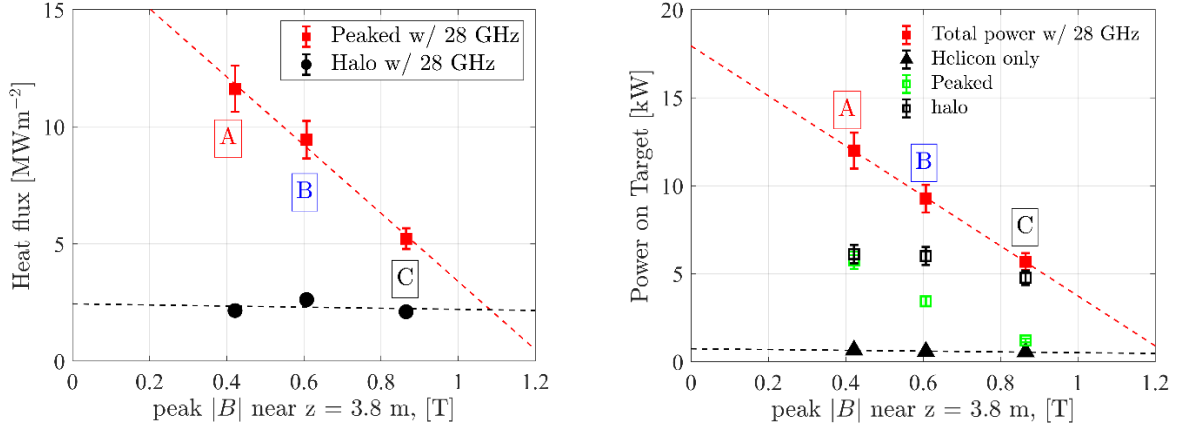

Figure 12, Heat flux (left) and total power (right) coupled to the target as a function of the peak magnetic field near z = 3.8 m. Heat flux is measured in two different regions: (black circles) "halo" and (red squares) "peaked" region observed in Figure 11. Measurements are labelled "A", "B" and "C" and are associated with the magnetic field profiles presented in Figure 10. The dotted lines represent linear fits to the data to demonstrate the linearity of the response and the slope of each data set.

### 5.1.2 Total power coupled to the target

Integrating the heat flux distributions $q(x_*, y_*, t)$ presented in Figure 11 over the entire surface area of the target plate $A$, as described in Eq. 4, provides the total power $\dot{Q}(t)$ transported to the target by the plasma. Figure 12 (right) presents the total power transported to the target during 28 GHz injection (red squares) and during 13.56 MHz only (black triangles) as a function of the peak magnetic field near z = 3.8 m. The green squares represent the power due to the "Peaked" heat flux and it obtained by integrating inside the green contour lines in Figure 11. The "Halo" term corresponds to the region outside the green contour lines. The experimental data demonstrates that for the same injected 28 GHz microwave power (70 kW), profile A is significantly more effective at transporting power to the target at a value of 12 kW, while profile C only delivers about 5.7 kW. This observation is true *only* during the application of the 28 GHz microwave power. The data in Figure 12 (right) indicates that the magnetic field profile has little effect on the power transport during helicon-only plasmas (black circles). The linear trend between coupled power and magnetic field profile during 28 GHz microwave injection has been previously observed experimentally, albeit with much heating performance about 10 times lower: 1.5 kW of power coupled to the target with 50 kW of microwave power [16]; moreover, this trend has also been observed numerically using linear Fokker-Planck calculations presented in reference [16] where 28 GHz microwaves are assumed to be resonantly absorbed at the 2nd harmonic resonance layer.

$$\dot{Q}(t) = \int q \, dA = \iint q(x_*, y_*, t) \, dx_* dy_* \qquad \text{Eq. 4}$$

### 5.1.3 Heating efficiency

The efficiency for the 28 GHz microwave heating system is described by Eq. 5, where $\dot{Q}_0$ represents the total power incident on the target with *only* 13.56 MHz power (helicon only) and $\dot{Q}_1$ with *both* 13.56 MHz and 28 GHz power sources. The term $P_{28}$ represents the injected 28 GHz power. Using the data presented in Figure 12 (right) the heating efficiency is presented in Figure 13. The data demonstrates that the heating efficiency reaches a maximum value of 17% and decreases to about 7% as the peak magnetic field near z = 3.8 m is increased. This is an important observation, since it demonstrates that heating efficiency can be controlled by the choice of the magnetic field profile. These results are a significant improvement over previously reported measurements; for example, in reference [16] the electron heating efficiency is calculated to be about 3% using Eq. 5.

$$\eta_{28} = \frac{\dot{Q}_1 - \dot{Q}_0}{P_{28}} \qquad \text{Eq. 5}$$

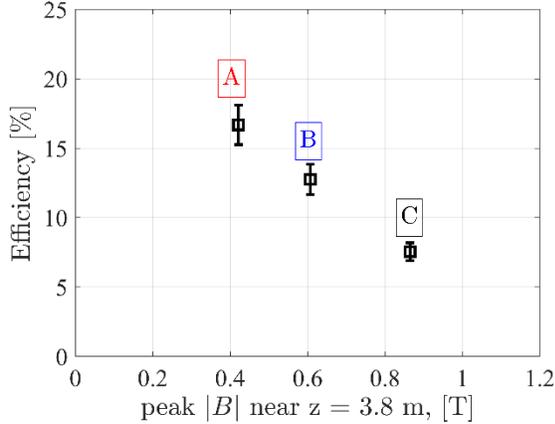

Figure 13, Electron heating efficiency based on data shown in Figure 12 (right) and using Eq. 5. The labels A to C represent the magnetic field profile (Figure 10) associated with the measurement.

## 5.2 Electron heating power scan

Using the same experimental setup described in the previous section and the magnetic field profile A (see Figure 10) to maximize parallel power transport to the target, the 28 GHz microwave power was systematically increased from 5 to 70 kW. The plasma-induced heat flux and the total power incident on the target plate was obtained using the IR thermographic diagnostic (section 2). The main plasma was produced with 500 ms pulses of 13.56 MHz RF power at approximately 120 kW and deuterium gas. The relative timing between the RF and the 28 GHz pulse is the same as that shown in Figure 9.

### 5.2.1 2-D Reconstructed Target Heat Fluxes

Figure 14 presents the two-dimensional (2-D) reconstructed heat flux distribution on the target with 70 kW of 28 GHz microwave power. The dotted circle represents the LUFS. The dashed straight line represents the path of a Double-Langmuir Probe (DLP) about 5 cm upstream of the tungsten target plate (Figure 17). The use of DLPs in RF environments has been discussed in references [18], [40]. The black square describes the region in space where the "halo" heat flux value was obtained. The green arrows illustrate the 28 GHz microwave injection zone. The green contour line represents the region where 50% of the integrated heat flux is found. The rest of the power is distributed in the "halo" region. The "peaked" heat flux contribution (hot spot) is again well correlated to the location of 28 GHz microwave injection. The heat flux distribution can be again described by a "peaked" and "halo" contribution.

Figure 15 (left) presents the "peaked" and "halo" heat flux contributions as a function 28 GHz microwave power. The data demonstrate that both contribution scale linearly with microwave power. The largest "peaked" and "halo" heat fluxes are approximately 22 $MWm^{-2}$ and 3 $MWm^{-2}$ respectively using 70 kW of 28 GHz microwave power. A peak heat flux of 22 $MWm^{-2}$ represents the highest heat flux measured in Proto-MPEX thus far with any combination of heating systems.

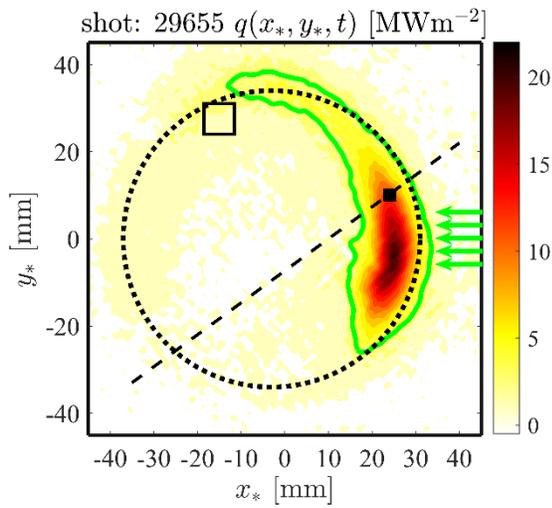

Figure 14, 2-D heat flux distribution at the target with the highest heat flux (22 MWm$^{-2}$) recorded in Proto-MPEX during electron heating experiments. Injected microwave power (28 GHz) was 70 kW. The dotted line represents the path of a Double Langmuir Probe (DLP) positioned 5 cm upstream of the target. The black solid square represents the location where the DLP has measured the highest electron temperature and is used to produce Figure 18. The green contour line represents the region containing 50% of the integrated heat flux.

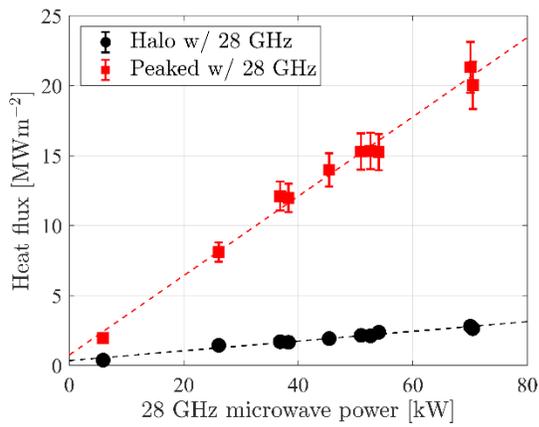
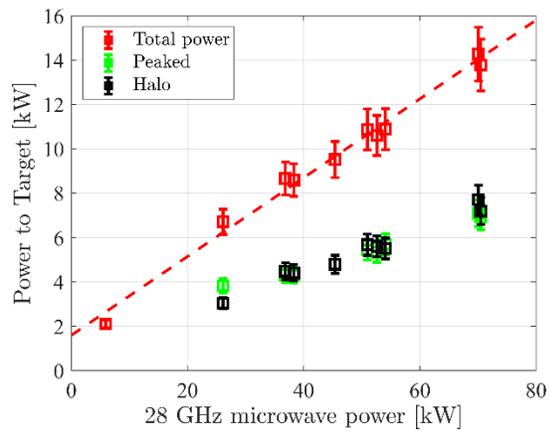

Figure 15, Heat flux (left) and total power (right) coupled to the target as a function of 28 GHz microwave power. The heat flux is measured in two different regions: (black circles) "halo" and (red squares) "peaked" region. The total power includes the contribution of both the 13.56 MHz and 28 GHz power. Lines represent best linear fits to the data.

### 5.2.2 Total power coupled to target and heating efficiency

The total power coupled to the target plate (Eq. 4) as a function of injected 28 GHz power is presented in Figure 15 (right). The "peaked" power represents the integrated heat flux inside the green contour line presented in Figure 14. The data demonstrates that the total heat load to the target scales linearly with applied heating power and delivers up to 14 kW when using 70 kW of 28 GHz microwave power. Moreover, the heating efficiency, based on the data from Figure 15 (right), is presented in Figure 16. Except for the first data point to the left in Figure 16, the 28 GHz heating efficiency is approximately constant at a value of about 18 % for this data set.

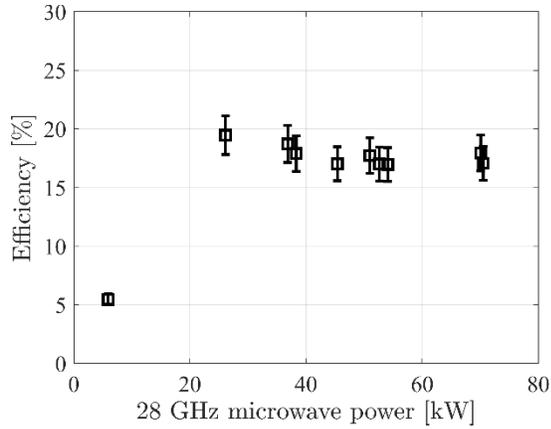

Figure 16, Electron heating efficiency based on data shown in Figure 15 (right) and using Eq. 5.

### 5.3 Plasma density and temperature measurements near the target plate

Figure 17 presents plasma density, electron temperature and electron pressure radial profiles measured with a Double Langmuir Probe (DLP) about 5 cm upstream of the target plate and along the dotted line presented in Figure 14. In Figure 17, the red/black traces represent the radial profiles with/without injection of 28 GHz microwave power. The relative timing between the 13.56 MHz and 28 GHz microwave pulse is presented in Figure 9.

The DLP measurements indicate that without 28 GHz power, the radial density profile has a triangular profile with a peak density of about $3.5 \times 10^{19}$ m$^{-3}$ and a typical hollow electron temperature profile in the range of 2-3 eV. Upon the application of 28 GHz microwaves (70 kW), the radial density profile becomes broader with the edge density increasing up to $2 \times 10^{19}$ m$^{-3}$; while the electron temperature indicates strong heating on right-side of the profile which corresponds to the region sampling the "hot spot" observed in Figure 14. Using both the density and temperature profiles, the electron pressure profile is calculated and indicates a significant increase in electron pressure in the plasma periphery with the largest contribution closest to the 28 GHz injection zone.

Positioning the DLP at the location of highest electron temperature (radius of 3.5 cm, black square in Figure 14), the 28 GHz microwave power was systematically increased. The resulting measurements are presented in Figure 18 where the red/black traces represent the measurements with/without 28 GHz injection. The electron density increases considerably as a function of microwave power while the electron temperature appears to have a weaker dependency on heating power and reaching values up to 9 eV. The electron pressure shows an almost linear dependency on heating power. This is also observed for the total power coupled to the target presented in Figure 15 (right). The electron temperature at the center of the "hot spot" is expected to be much higher than the measured values in Figure 18; however, diagnostic access prevented such measurement. Moreover, it is likely that a direct measurement of the "hot spot" with a DLP would lead to probe damage due to the intense heat flux (< 22 MWm$^{-2}$).

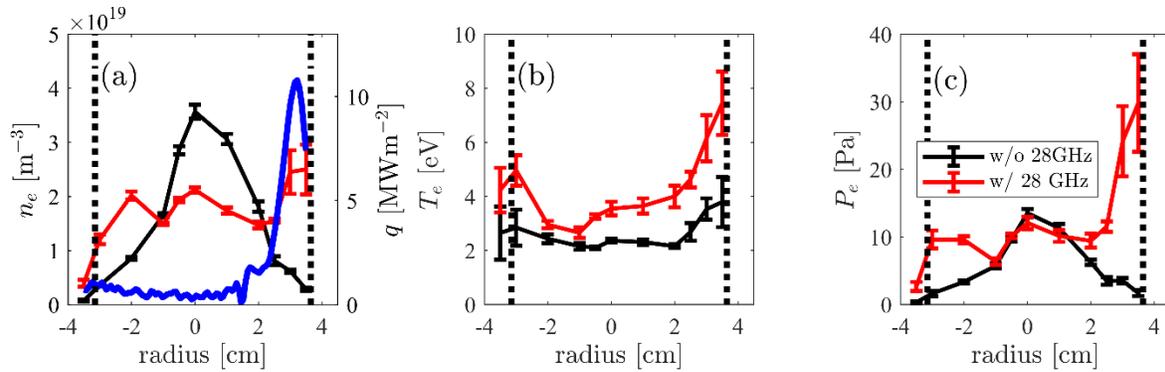

Figure 17, Radial (a) electron density, (b) electron temperature and (c) electron pressure profiles measured with a Double Langmuir Probe (DLP) 5 cm upstream of the tungsten target along the path shown in Figure 14. The dotted vertical lines represent the LUFS. The blue trace in (a) represents the IR-based heat flux on the target along the path shown in Figure 14.

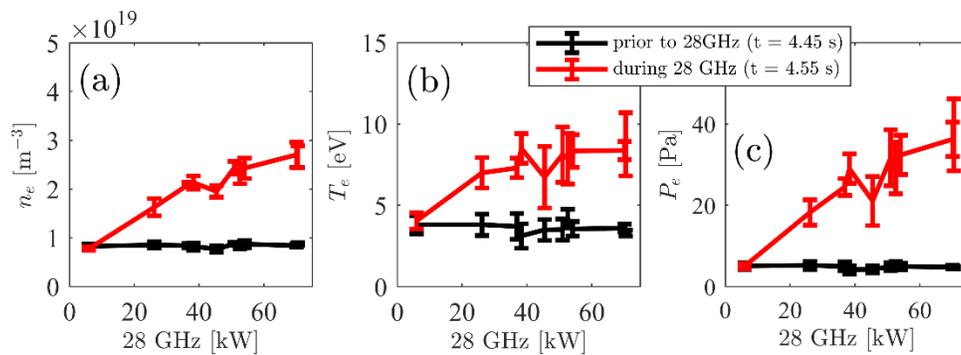

Figure 18, (a) Electron density, (b) electron temperature and (c) electron pressure as a function of 28 GHz microwave power. Data obtained with a Double Langmuir Probe (DLP) located 5 cm upstream of the tungsten target and close to the hot spot at a radial position of 3.5 cm. Black traces represent measurements taken prior to the 28 GHz pulse (see Figure 9).

# 6 Numerical results

Ray-tracing calculations have been performed to model the propagation and absorption of 28 GHz microwaves injected in Proto-MPEX. The GENRAY-C code is used [9] and the conditions are based on the experimental setup presented in section 2.2.

It is worth noting that the plasma density and electron temperature in the electron heating chamber were not measured during the present experiments. Fortunately, these measurements have been performed under equivalent experimental conditions [10, Fig. 5]. These measurements indicate that the peak plasma density in the electron heating chamber is over-dense with the O-mode cutoff at the periphery. Therefore, the plasma density and temperature in the present ray-tracing calculations are chosen to be consistent with the experimental observations from reference [10, Fig. 5].

The simulation setup is the following: (1) The magnetic field profile "A" (Figure 10) is used. (2) Operating frequency is set to 28 GHz. (3) Microwave beam is launched from a radius of r = 0.1 m and z = 3.23 m. (4) Microwave beam is injected at a 30-degree angle of incidence with an O-mode polarization. (5) Cyclotron absorption and Coulomb collisions are included. (6) The radial density profile in the electron heating chamber is assumed to be over-dense with a peak value of $2 \times 10^{19}$ m$^{-3}$ (Figure 20 (right)). (7) Edge electron density is set to $1 \times 10^{18}$ [m$^{-3}$] all the way to the wall. (8) Electron temperature profile is assumed to be uniform at a value of 4 eV.

The main results are presented in Figure 19 to Figure 21. Figure 19 illustrates the on-axis magnetic field strength in the electron heating chamber and the 2-D trajectory of the microwave beam as it propagates into the plasma column. For this magnetic field scenario, there are several harmonic electron cyclotron resonance surfaces. In fact, the microwave beam travels through a 3$^{rd}$ harmonic resonant surface. The ray-tracing calculation indicates that under these conditions, 3$^{rd}$ harmonic cyclotron O-mode absorption of the ray is negligible. This is likely due to the low electron temperature (4 eV), which leads to low single-pass absorption. However, as the microwave beam reaches the Upper Hybrid Resonance (UHR) layer, it is promptly absorbed. All the absorption occurs within the region bounded by the black rectangle. A close-up view of this absorption region is presented in Figure 20 to illustrate where the power is absorbed.

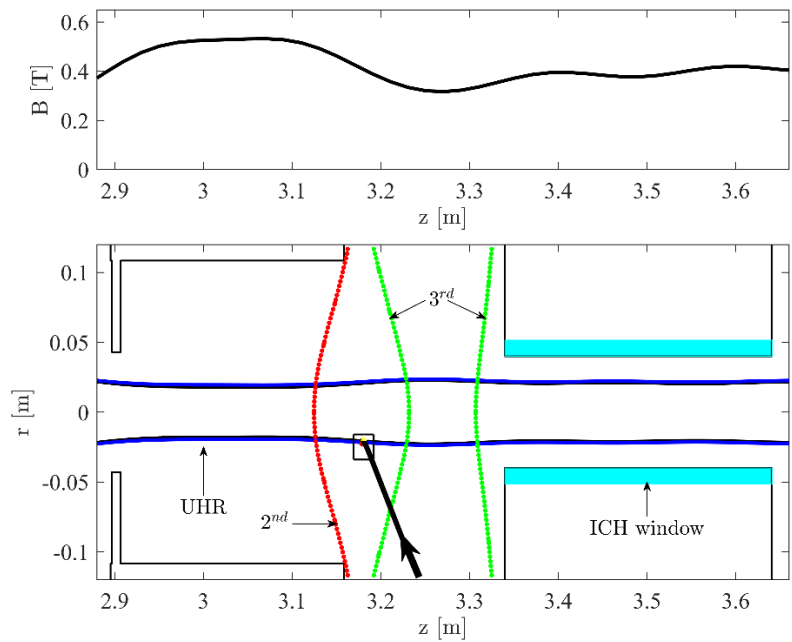

Figure 19, (top) On-axis magnetic field profile throughout the electron heating chamber. (bottom) schematic describing the ray-tracing calculation setup. Microwave beam is injected from the bottom at an incidence angle of 30 degrees. Red and green lines represent 2$^{nd}$ and 3$^{rd}$ harmonic resonance layers. The small rectangle encloses the region where strong absorption is observed (see Figure 20).

A close-up view of the ray at the absorption region is presented in Figure 20 (left) where the UHR and O-mode cutoff layers can be observed. As the microwave beam reaches the O-mode cut-off layer, it changes direction until the ray nears the UHR layer. The ray is then promptly absorbed. The radial absorption profiles due to both cyclotron (resonant) and collisional (non-resonant) interactions are presented in Figure 20 (right). These profiles indicate that the microwave beam is absorbed via *collisions* just outside the UHR layer and *resonantly* in the over-dense region near the O-mode cutoff. Further inspection of the ray-tracing calculations indicate that the *resonant* absorption is caused by doppler-shifted 2$^{nd}$ harmonic electron cyclotron resonance of the Bernstein (B) mode. The doppler shift is caused by the strong increase in the parallel refractive index of the B-mode up to values of 25. Calculations indicate that most of the microwave power is absorbed in the under-dense region and a small fraction in the over-dense region.

To gain further insight into the absorption mechanism, Figure 21 presents the power absorption of the ray along with the perpendicular refractive index as a function of the ratio $\omega_{pe}/\omega$. Comparing these results with those presented in *Appendix 3: Cold-plasma wave dispersion and O-X-B mode conversion*, the wave branches present in the ray-tracing calculation are identified: O-mode, slow X-mode and Bernstein mode. By presenting the data in this format (Figure 21), two important physics become evident: (1) the O-mode wave undergoes O-X conversion upon reaching the O-mode cutoff and (2) absorption is *entirely* due to the excitation of the B-mode via X-B mode-conversion at the UHR. Absorption continues as the B-mode propagates into the over-dense region where both cyclotron (resonant) and collisional absorption is observed.

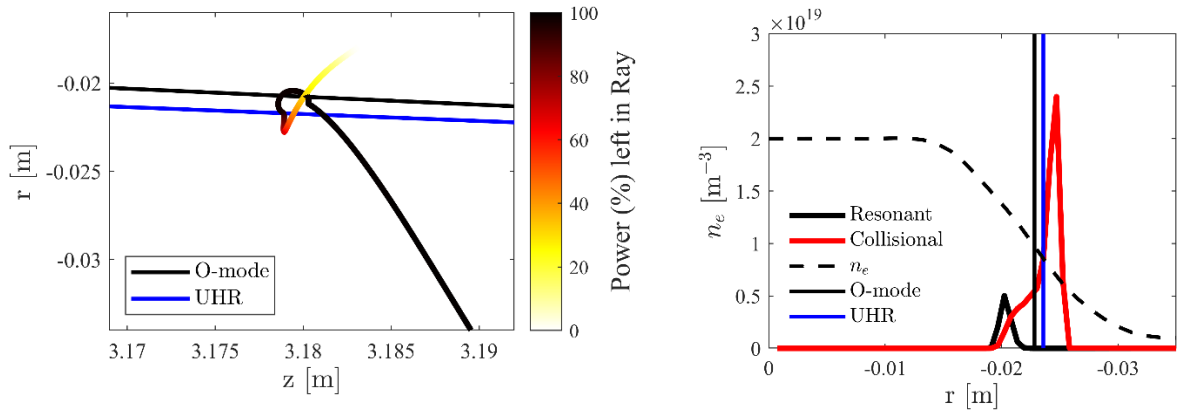

Figure 20, (left) close-up view illustrating the behavior of the ray near the Upper-Hybrid Resonance (UHR) layer and O-mode cutoff. (right) Radial profiles of plasma density and microwave absorption profiles in the electron heating chamber. Both resonant and non-resonant (collisional) absorption profiles are presented. The vertical lines represent the UHR and O-mode cutoff layers. Electron temperature is uniform with a value of 4 eV.

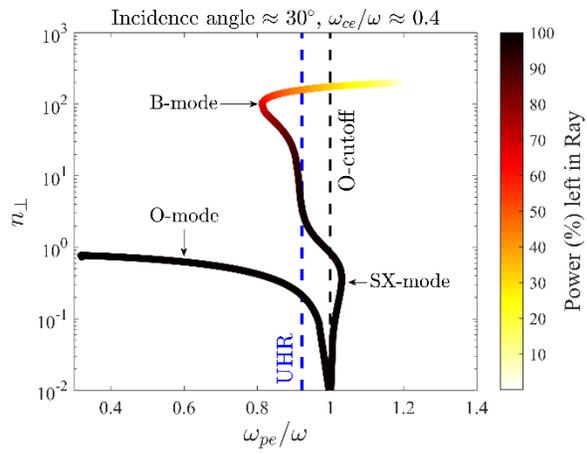

Figure 21, dispersion relation obtained from the ray-tracing calculation indicating the power absorption strength of the various wave branches. The microwave beam is injected with O-mode polarization. At that point, no absorption has occurred, and the power left on the beam is 100 %. The beam starts to lose power once it crosses the UHR layer as a SX-mode and the power left on the beam reaches 0% along the B-mode branch.

# 7 Discussion
## 7.1 Heat flux measurements

The experimental results demonstrate two important capabilities: (1) effective power coupling to the target using oblique microwave injection, and (2) effective control of power transport to the target by varying the magnetic field.

The data presented in Figure 12 clearly demonstrates that both the peak heat flux and power coupled to the target can be controlled by varying the magnetic field downstream of the electron heating chamber. These observations have contributed to the development of MPEX by guiding the design of the magnetic field profile and demonstrating a control variable to maximize power transport and heating efficiency. Based on the present experimental data, the mechanism enabling this type of control cannot be determined. Numerical results presented in reference [16] explore this process and discuss possible mechanisms related to the transport of fast electrons created during $2^{nd}$ harmonic resonance.

By using the optimal magnetic field (profile "A") up to 20% of microwave power is accounted for at the target via plasma transport (Figure 16). In fact, most of that power is concentrated in a small region at the plasma edge. The recorded peak heat fluxes of ~20 $MWm^{-2}$ and heating efficiencies up to 17-20 % represents a significant improvement over any Proto-MPEX results previously published. A more complete measure of the efficiency can be obtained by diagnosing both the target and dump plates.; however, capability was not available for the present experiments. Moreover, radiated power measurements were not available due to the lack of diagnostics. For this reason, a detailed audit of the power balance has not been provided.

An important observation is the concentration of heat flux at the plasma edge directly at the microwave injection zone (Figure 14). Previous experiments [17] with 28 GHz oblique microwave injection from the *top* of the device show that the peak heat flux does in fact follow the injection zone. An example of the target heat flux profile measured during such experiments is presented in Figure 22. These observations demonstrate that the location of the microwave launcher controls the azimuthal location of the peak heat flux. The interpretation is that power is absorbed locally at the microwave injection zone and transported along magnetic flux lines from the heating region to the target. This is further supported by Figure 11 where the radial location of the peak heat flux follows the LUFS as the magnetic field at the target is varied. In other words, there is a direct mapping between the peak heat flux at the target and the microwave absorption zone. Based on these considerations, multiple launchers may be positioned around the plasma column to reduce the azimuthal asymmetry of the heat flux at the target.

The strong localization of the heat flux at the target is not optimal for the exposure of large samples for PMI studies. A broad and uniform heat flux pattern is required such as those presented in reference [11], albeit at much higher densities and heat fluxes. The strategy to optimize the heat flux pattern for PMI studies depends on the electron heating mechanisms at play. The electron heating mechanism is discussed in the next section based on ray-tracing calculations and experimental observations from the present and previous work.

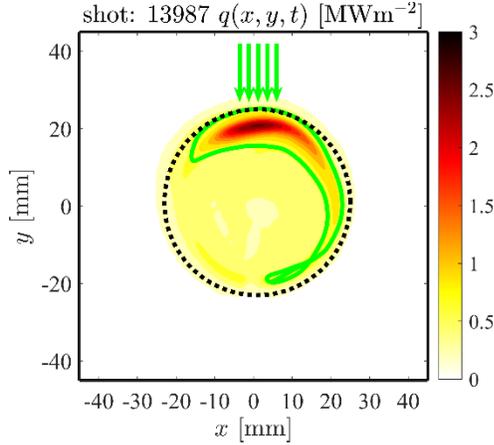

Figure 22, 2-D heat flux pattern at the target measured during 28 GHz electron heating experiments with oblique microwave injection. The green arrows indicate the location of the microwave injection. Experimental setup is described in reference [17]. The green contour line encloses about 1/3 of integrated heat flux.

## 7.2 Ray-tracing calculations

The ray tracing shows that the microwave beam is absorbed in a single pass at the plasma edge (Figure 20) via an O-X-B process (Figure 21). The edge absorption is consistent with the experimental observation that most of the microwave power coupled to the plasma is concentrated at the plasma edge directly at the beam injection zone (Figure 14, Figure 22). Moreover, ray tracing shows no observable absorption of the microwave beam as it travels through a 3rd harmonic resonance (Figure 19). Inspection of the ray tracing results confirms that the O-mode 3rd harmonic (O3) absorption is exceedingly small.

Given the launch geometry, the polarization of the beam and the presence of a 2nd harmonic resonance, we must consider the absorption of the O-mode at the 2nd harmonic resonance (O2). For completeness, we also consider X-mode 2nd harmonic (X2). The O2 and X2 single-pass absorption can be calculated using Eq. 6, where $P_{abs}$ is the absorbed power, $P_0$ is the injected power and $\tau$ is the so-called "optical depth" [41]. The optical depths for oblique injection are taken from reference [42, pp. 1206, Table 12]. Using an angle of incidence of 30 degrees, the magnetic field from the present experiment and a density of $1 \times 10^{18}$ [m$^{-3}$] to allow the 28 GHz beam to propagate through the plasma, the single-pass absorption for O2 and X2 oblique injection is presented in Figure 23 for a range of electron temperatures (1 eV to 10 keV).

Figure 23 indicates that at the electron temperatures found in the present experiment (< 10 eV), the O2 and X2 single-pass absorption is exceedingly small (<10$^{-2}$). Multi-pass absorption in the cavity formed by the electron heating chamber would lead to a symmetric and distributed absorption profile. Only at fusion relevant temperatures (>1 keV) is strong single-pass absorption expected. This means O2, O3 or X2 cyclotron absorption are unlikely mechanisms for the microwave absorption in the present experiment.

$$\frac{P_{abs}}{P_0} = (1 - \exp(-\tau)) \qquad \text{Eq. 6}$$

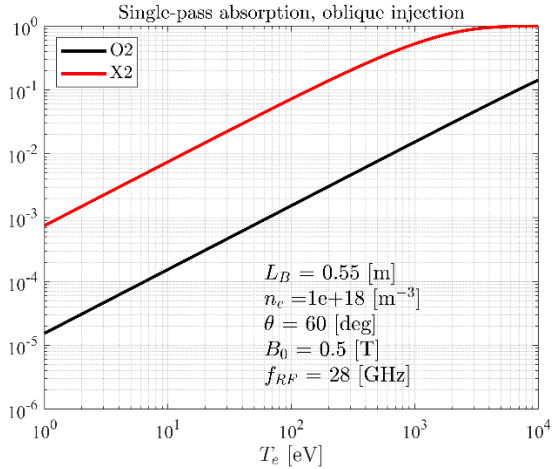

Figure 23, Single-pass absorption for O2 and X2 with oblique propagation relative to the magnetic field. The conditions for the calculation are shown in the figure. The magnetic field gradient scale length is given by $L_B = B(dB/ds)^{-1}$. The term $\theta$ represents the angle between the ray and the magnetic field.

Therefore, based on the measured over-dense conditions in the electron heating region [10, Fig. 5], the edge absorption predicted by the ray tracing, the O2,X2 single-pass absorption calculations at < 10 eV and the observed heat flux localization at the injection zone, the most likely mechanism responsible for the microwave absorption in the present experiment is O-X-B at the plasma edge. Moreover, ray-tracing shows that *all* the absorption is due to the B-modes in the *under-dense* and *over-dense* regions. More specifically, absorption is collisional in the *under-dense* region while a small fraction is absorbed both collisionally and resonantly in the *over-dense* region next to the O-mode cutoff (Figure 20). Interestingly, ray-tracing indicates that resonant absorption in the over-dense region is due to doppler-shifted 2<sup>nd</sup> harmonic electron resonance of the electron Bernstein wave.

### 7.3    Plasma density and temperature measurements at the target

Radial density and temperature profiles have been measured 5 cm upstream of the target (Figure 17). The profiles undergo significant modifications upon application of the microwave power. The mechanisms driving these changes are not fully understood. Potential mechanisms are presented supported by either measurements or numerical results.

The electron temperature reaches a maximum value of 8 eV at the plasma edge closest to the microwave injection zone. This is consistent with the location of the peak heat flux (Figure 14); Moreover, a modest temperature increase is observed over the entire diameter. This is suggestive of electron heating in the core plasma; however, this is not consistent with the ray-tracing calculations which show edge absorption; albeit some of it is absorbed resonantly in the over-dense plasma.

One possible mechanism for the observed electron heating in the core is increased penetration depth of the B-mode caused by localized heating at the injection zone. This process is not captured by the ray-tracing code since the electron temperature is held constant. Reference [9] discusses the effect of collisions on the O-X-B process and shows that the penetration depth of B-modes is significantly reduced with collisions. Since Coulomb collisions scale strongly with temperature ($\propto n_e T_e^{-3/2}$), it is entirely possible that localized heating increases the electron temperature sampled by the wave; thereby, reducing collisionality and enabling better penetration of the B-mode into the over-dense region. This is supported by the numerical results presented in reference [9, Fig. 12].

The electron density drops considerably in the center while increasing at the edge (Figure 17). From a PMI point of view, this drop is undesirable since high plasma densities are required. The mechanism behind this process is not known and is currently under investigation. It is worth noting that plasma flow measurements with Mach probes presented in reference [43, Fig. 6.14] indicate strong plasma acceleration occurs in the target region during 28 GHz microwave heating experiments. The parallel flow velocity increases to about twice its value prior to the application of microwave power. If the total particle flux is conserved, the observed acceleration is consistent with the halving of the central density observed experimentally (Figure 17). Additional experimental evidence in combination with numerical work is required to make this conclusive.

At the location of the heat flux maxima, the plasma density increases almost linearly with microwave power (Figure 18). The electron temperature, however, appears to saturate at about 8-10 eV. This type of scaling is also observed in plasma sources where density increases linearly with applied power while the electron temperature is fixed provided enough neutral gas is provided. Considering that an electron temperature of 8-10 eV is sufficient for ionizing deuterium gas and given the high neutral gas pressures observed in the target region (~0.5–1 Pa) due to plasma-surface recombination [5], it is quite possible that the increase in edge density is caused by neutral gas ionization. The energy source is the power transported to the target during microwave heating. This explanation can be tested by puffing gas directly at the target region and observing the changes in density and temperature.

### 7.4 Centrally peaked microwave absorption

If we consider that the most likely microwave absorption process is O-X-B, the radial absorption profile is related to the position of the UHR layer. Given the dependence of the UHR layer on plasma density and magnetic field (Eq. 2), the microwave absorption layer in an inhomogeneous plasma column can be controlled by varying: (1) the magnetic field, (2) microwave frequency and/or (3) plasma density. This has been demonstrated experimentally in references [8], [11]. In the present experiments, the microwave absorption can be shifted to the core by increasing the microwave frequency while keeping the ratio $\omega_{ce}/\omega$ constant to keep the location of the 2$^{nd}$ harmonic electron cyclotron resonant surface unchanged.

In the context of MPEX, a microwave frequency of 70 GHz and a magnetic field of 1.25 Tesla has been chosen. This keeps the ratio $\omega_{ce}/\omega \approx 0.5$ and moves the UHR layer to a density of about $4.5 \times 10^{19}$ m$^{-3}$ and the O-mode cut-off density to about $6 \times 10^{19}$ m$^{-3}$. This is well within the plasma density values expected in the electron heating region of MPEX using the planned 200 kW 13.56 MHz RF transmitter. The RF power can be adjusted to position the UH resonance layer in the plasma core. Moreover, the microwave injection system will use ellipsoidal reflectors to focus the microwaves into the plasma to favor coupling. The magnetic field profile has been chosen to maximize the power transport to the target as observed in Figure 12.

## Acknowledgements

This material is based upon work supported by the U.S. Department of Energy, Office of Science, Office of Fusion Energy Sciences, under Contract No. DEAC05-00OR22725.

# Appendix 1: Forward problem

The geometry of the problem is illustrated in Figure 24 where $L_x, L_y$ and $L_z$ are the spatial dimensions of the target plate. The goal is to solve for the time-dependent temperature distribution $T$ given a set of boundary condition and an initial temperature distribution. The heat diffusion equation with *constant* material properties is given by Eq. 7. The initial condition (IC) and the boundary conditions (BC) are given in Eq. 8 and Eq. 9 respectively. Spatial and temporal coordinates are given by $\bar{x}, \bar{y}, \bar{z}$ and $\bar{t}$ respectively. The thermal diffusivity $\alpha$ is given by $\alpha = k/\rho c_p$, where $k$ is the thermal conductivity, $\rho$ the mass density and $c_p$ the heat capacity. The term $q$ represents the surface heat flux which is in general a function of space and time.

$$\frac{\partial T}{\partial \bar{t}} = \alpha \left( \frac{\partial^2 T}{\partial \bar{x}^2} + \frac{\partial^2 T}{\partial \bar{y}^2} + \frac{\partial^2 T}{\partial \bar{z}^2} \right) \qquad \text{Eq. 7}$$

$$T(\bar{x}, \bar{y}, \bar{z}, \bar{t} = 0) = T_0(\bar{x}, \bar{y}, \bar{z}) \qquad \text{Eq. 8}$$

$$\left.\frac{\partial T}{\partial \bar{x}}\right|_{\bar{x}=0} = 0 \quad \left.\frac{\partial T}{\partial \bar{x}}\right|_{\bar{x}=L_x} = 0$$

$$\left.\frac{\partial T}{\partial \bar{y}}\right|_{\bar{y}=0} = 0 \quad \left.\frac{\partial T}{\partial \bar{y}}\right|_{\bar{y}=L_y} = 0 \qquad \text{Eq. 9}$$

$$\left.\frac{\partial T}{\partial \bar{z}}\right|_{\bar{z}=0} = -\frac{q}{k} \quad \left.\frac{\partial T}{\partial \bar{z}}\right|_{\bar{z}=L_z} = 0$$

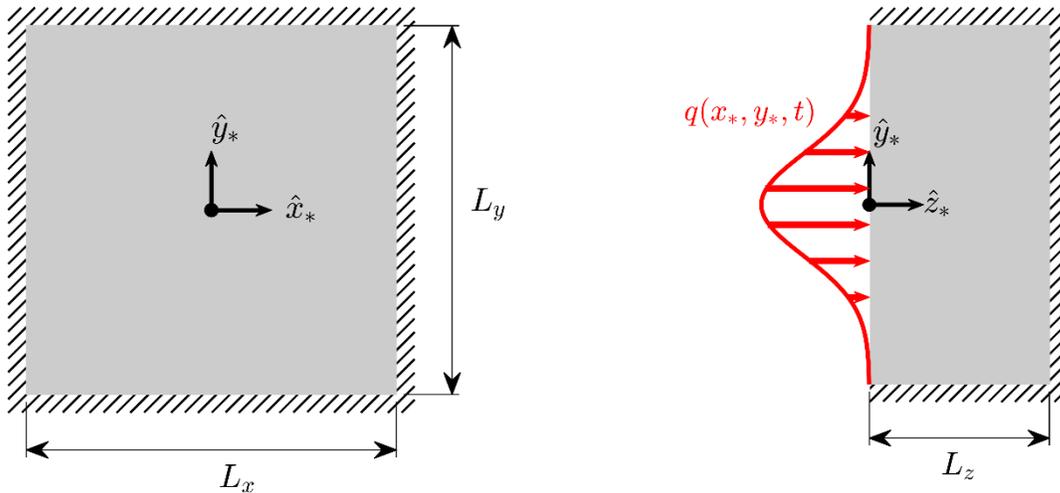

Figure 24, (left) front and (right) side view of the target plate geometry used to solve the heat diffusion problem. All surfaces have insulating boundary conditions except for the "front-side" surface which has an applied surface heat flux distribution $q(x_*, y_*, t)$.

The normalized version of Eq. 7 is given by Eq. 10 where $x, y, z$ and $t$ represent the *normalized* spatial and temporal coordinates given by Eq. 11 and $L$ represents the characteristic length of the geometry. The *normalized* temperature distribution is given by $u$ in Eq. 12 where $T^*$ represents the characteristic temperature given by Eq. 13 and $q^*$ represents the characteristic heat flux of the problem. Moreover, the *normalized* surface heat flux $h$ is given by Eq. 14.

$$\frac{\partial u}{\partial t} = \frac{\partial^2 u}{\partial x^2} + \frac{\partial^2 u}{\partial y^2} + \frac{\partial^2 u}{\partial z^2} \qquad \text{Eq. 10}$$

$$x = \bar{x}/L \quad y = \bar{y}/L \quad z = \bar{z}/L \quad t = \bar{t}\alpha/L^2 \qquad \text{Eq. 11}$$

$$u = \frac{T - T_0}{T^*} \qquad \text{Eq. 12}$$

$$T^* = \frac{q^* L}{k} \qquad \text{Eq. 13}$$

$$h = \frac{q}{q^*} \qquad \text{Eq. 14}$$

The solution to Eq. 10 subject to the IC and BCs presented in Eq. 8 and Eq. 9 is given by Eq. 15 where $U_{mn}$ is given by the convolution integral in Eq. 16 and represents the Fourier-cosine components of the solution $u$. The terms $p_m(x)$ and $p_n(y)$ represent the eigenfunctions of the problem and are given by Eq. 17. The terms $\beta_x, \beta_y$ and $\beta_z$ are given by Eq. 18. The term $H_{mn}$ is given in Eq. 19 and represents the Fourier-cosine components of the surface heat flux $h$. The terms $\delta_{mm}$ and $\delta_{nn}$ are given by Eq. 20. The term $K_{mn}$ is a special function given by Eq. 21 where $G(z, t)$ represents the 1D Green's function of the problem [20] given in Eq. 22 and the terms $\lambda_{mn}^2$ and $\lambda_l^2$ are given in Eq. 23.

$$u(x, y, z, t) = \sum_{n=0}^{\infty} \sum_{m=0}^{\infty} U_{mn}(z, t) p_m(x) p_n(y) \qquad \text{Eq. 15}$$

$$U_{mn}(z, t) = \frac{1}{\beta_z} \int_0^t H_{mn}(t') K_{mn}(z, t - t') dt' \qquad \text{Eq. 16}$$

$$p_m(x) = \cos\left(m\pi \frac{x}{\beta_x}\right) \quad p_n(y) = \cos\left(n\pi \frac{y}{\beta_y}\right) \quad p_l(z) = \cos\left(l\pi \frac{z}{\beta_z}\right) \qquad \text{Eq. 17}$$

$$\beta_x = \frac{L_x}{L} \quad \beta_y = \frac{L_y}{L} \quad \beta_z = \frac{L_z}{L} \qquad \text{Eq. 18}$$

Eq. 19

$$H_{mn}(t) = \frac{1}{\beta_x \beta_y} \frac{1}{\delta_{mm}\delta_{nn}} \int_0^{\beta_y} \int_0^{\beta_x} h(x,y,t) p_m(x) p_n(y) \, dx \, dy$$

$$\delta_{ij} = \begin{cases} 0 & i \neq j \\ \frac{1}{2} + \frac{1}{2}\text{sinc}((i+j)\pi) & i = j \end{cases} \qquad \text{Eq. 20}$$

$$K_{mn}(z,t) = \exp(-\lambda_{mn}^2 t) \, G(z,t) \qquad \text{Eq. 21}$$

$$G(z,t) = \sum_{l=0}^{\infty} \left[ \frac{\exp(-\lambda_l^2 t)}{\delta_{ll}} p_l(z) \right] \qquad \text{Eq. 22}$$

$$\lambda_{mn}^2 = \left(\frac{m\pi}{\beta_x}\right)^2 + \left(\frac{n\pi}{\beta_y}\right)^2 \qquad \lambda_l^2 = \left(\frac{l\pi}{\beta_z}\right)^2 \qquad \text{Eq. 23}$$

The expression in Eq. 15 can be readily evaluated by truncating the infinite sum at values of $m$ and $n$ large enough to resolve the spatial features of both the heat flux and temperature distributions. For the present work, these values range between 40 to 60 mode numbers. Moreover, the infinite sum in Eq. 22 can be readily evaluated with a few hundred terms.

# Appendix 2: Validating the solution to the inverse problem

In section 3, the procedure for reconstructing the surface heat flux on the tungsten target via an inverse method is introduced. In this section, the validation process for the inverse solution is presented to demonstrate the applicability to the present work.

For the validation process, a tungsten target plate with the geometry presented in Figure 24 is chosen. The squared-shaped plate is chosen to be 90 mm in height as in the experimental setup in Proto-MPEX. The plate thickness is selected to be 0.75, 1.75, 3 and 4 mm. The synthetic heat flux distribution $q(x_*, y_*, t)$ on the "front-side" surface is described with a gaussian distribution in space $(x_*, y_*)$ and a gaussian variation in time. Based on the thickness of the plate, different peak heat fluxes of 5, 12, 20 and 27 MWm$^{-2}$ are used. Figure 25 compares the *synthetic* peak heat fluxes (black lines) against the *reconstructed* peak heat fluxes (red lines) obtained using the inverse solution for different plate thicknesses. These results show that as the plate thickness increases, the error in the *reconstructed* heat flux temporal evolution increases. These results demonstrate that for the cases with 0.75 mm and 1.75 mm thickness, the heat flux reconstruction is very good. In the experiments reported in this paper, the target plate has a thickness of 0.75 mm; hence, the error in the reconstruction is acceptably small as shown in Figure 25a. As the thickness increases beyond 1.75 mm, the error in the reconstruction becomes significant. The inverse heat flux analysis presented in the results section of this paper (section 5) is carried out using the same inverse method utilized in making Figure 25.

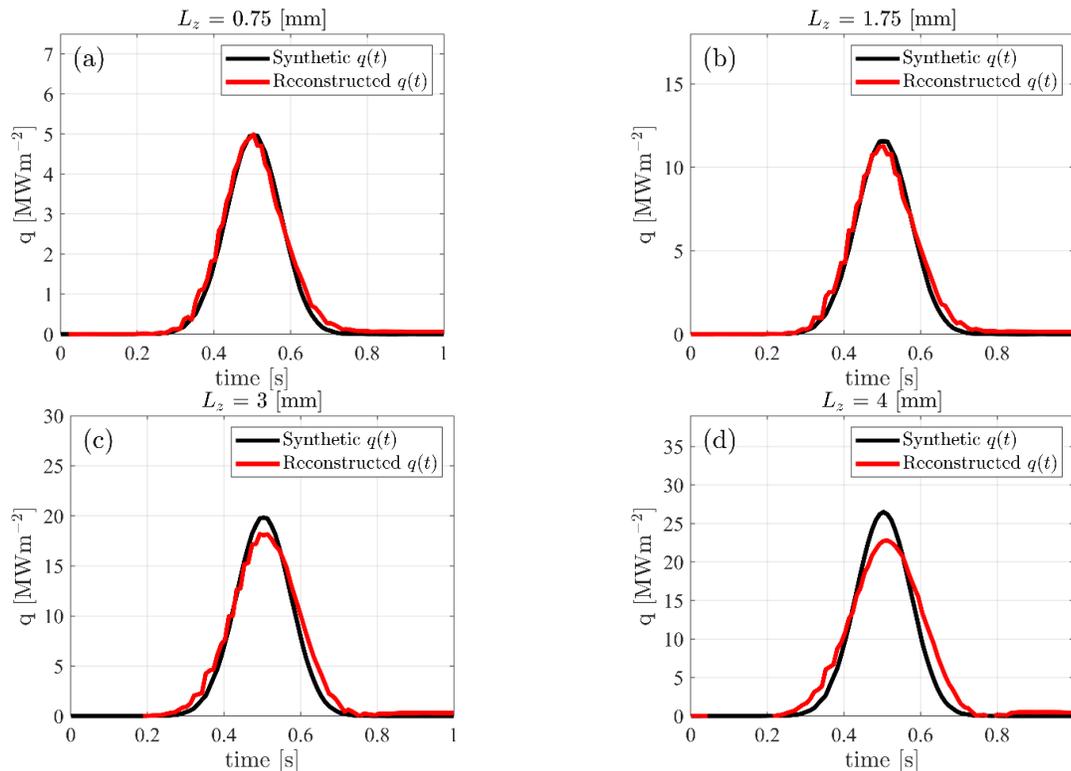

Figure 25, Comparing the time-evolution of the synthetic and reconstructed peak heat fluxes on a tungsten plate 90 mm by 90 mm for various plate thicknesses: (a) 0.75 mm, (b) 1.75 mm, (c) 3 mm and (d) 4 mm. The reconstructed heat flux was obtained by solving the inverse problem with constant material properties.

# Appendix 3: Cold-plasma wave dispersion and O-X-B mode conversion

The dispersion of cold-plasma waves in a uniform plasma is given by the well-known biquadratic equation in Eq. 24. Assuming that the frequency is high enough such that ion motion can be neglected, the Stix terms S, P, R and L are given in Eq. 25. The cold-plasma dispersion relation can be applied in slowly-varying plasma conditions with length scales much greater than the wavelengths under consideration.

$$Sn_\perp^4 - \left(RL + RS - n_\parallel^2(P + S)\right)n_\perp^2 + P(n_\parallel^2 - R)(n_\parallel^2 - L) = 0 \qquad \text{Eq. 24}$$

$$S = 1 - \frac{\omega_{pe}^2}{\omega^2 - \omega_{ce}^2}$$

$$P = 1 - \left(\frac{\omega_{pe}}{\omega}\right)^2$$

$$R = 1 - \left(\frac{\omega_{pe}}{\omega}\right)^2 \left(\frac{\omega}{\omega - \omega_{ce}}\right)$$

$$L = 1 - \left(\frac{\omega_{pe}}{\omega}\right)^2 \left(\frac{\omega}{\omega + \omega_{ce}}\right)$$

Eq. 25

Figure 26 (left) illustrates the dispersion for purely perpendicular propagation which corresponds to an angle of incidence of zero or $n_\parallel = 0$. A value of $\omega_{ce}/\omega = 0.52$ is used. Two polarizations propagate at normal angle of incidence: (1) O-mode and (2) X-mode. The O-mode has a cutoff labelled "O-cutoff" where $n_\perp = 0$ and corresponds to a zone of total wave reflection. The X-mode has two cutoffs labelled "FX-cutoff" and "SX-cutoff" which denote the slow and fast X-mode cutoff zones. In addition, the SX-mode has a resonance at the Upper Hybrid Resonance (UHR). At this location, the SX-mode becomes electrostatic, and its perpendicular refractive index increases without bound. Correct treatment of this resonance requires the use of hot plasma effects and leads to an additional branch: the electron Bernstein wave (B-mode) with a dispersion relation given by Eq. 26, where the ion motion has been neglected. In Eq. 26, the terms $k_\perp$ represents the perpendicular wavenumber and $r_{Le}$ the electron gyro-radius. At the UHR, the SX-mode has a confluence with the B-mode and mode conversion can occur. At normal angles of incidence, the B-mode can be excited by launching a FX-mode which tunnels through both the FX-mode cutoff (FX-cutoff) and UHR layer and reaches the SX-mode branch where it can connect to the B-mode.

$$1 - \frac{\omega_{pe}^2}{\omega}\frac{e^{-\lambda}}{\lambda} \sum_{n=-\infty}^{\infty} \frac{n^2 I_n(\lambda)}{\omega - n\omega_{ce}} = 0 \quad \text{where} \quad \lambda = \frac{1}{2}(k_\perp r_{Le})^2 \qquad \text{Eq. 26}$$

To improve coupling efficiency, the FX-mode tunnelling process can be completely removed by oblique O-mode injection [37]. At an incidence angle given by Eq. 27 below, the O-mode cutoff and the SX-mode cutoff become coincident and thus effective mode conversion from the O-mode to the SX-mode is possible as shown in Figure 26 (right). Moreover, provided an UHR layer is nearby, the SX mode can mode convert to a B-mode which propagates into the high density region without any cutoff density. Hence, injecting an O-mode polarized wave at an oblique angle of incidence into a radially-varying plasma column with an O-mode cutoff can lead to the transformation of an externally injected *electromagnetic* wave into a radially propagating *electrostatic* wave. This newly excited wave is readily absorbed *non-resonantly* via collisions and/or *resonantly* at nearby doppler-shifted harmonic cyclotron resonant surfaces [39].

$$\sin\theta = \sqrt{\frac{Y}{1+Y}} \quad \text{where } Y = \frac{\omega_{ce}}{\omega}$$

Eq. 27

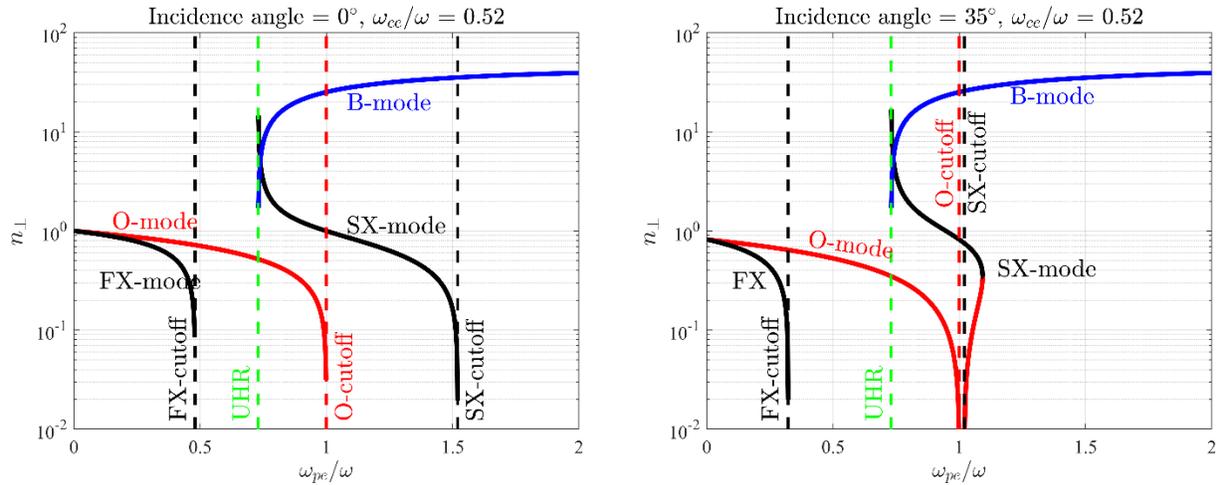

Figure 26, Cold-plasma wave dispersion for (left) perpendicular propagation (normal angle of incidence) and (right) oblique angle of incidence of 35 degrees. The blue line corresponds to the electron Bernstein wave (B-mode) and is calculated by solving the roots of Eq. 26. A ratio of $\omega_{ce}/\omega = 0.52$ is used.

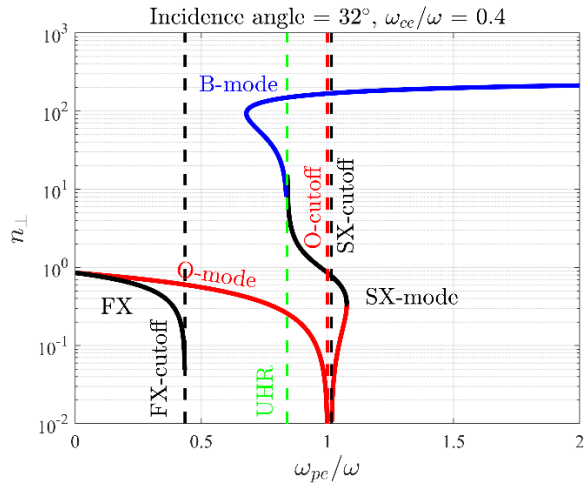

Figure 27, Same as Figure 26 (left) but with a ratio of $\omega_{ce}/\omega = 0.4$. Notice the different behavior of the B-mode.